\documentclass[aps,prm,amsmath,amssymb,reprint,superscriptaddress,nofootinbib]{revtex4-2}
\usepackage{charter,graphicx,verbatim,threeparttable,float,amssymb,gensymb }
\usepackage{upgreek}
\usepackage{booktabs}
\usepackage{tabularx}

\newcommand{\kagSS}{\textit{Ln}Ti$_3$(Sb,Sn)$_4$}
\newcommand{\SmSS}{SmTi$_3$(Sb,Sn)$_4$}

\newcommand{\ef}{E$_\text{F}$}
\newcommand{\NdSS}{NdTi$_3$(Sb,Sn)$_4$}
\newcommand{\GdSS}{GdTi$_3$(Sb,Sn)$_4$}
\newcommand{\PrSS}{PrTi$_3$(Sb,Sn)$_4$}
\newcommand{\CeSS}{CeTi$_3$(Sb,Sn)$_4$}
\newcommand{\LTB}{\textit{Ln}Ti$_3$Bi$_4$}

\begin{document}

\preprint{APS/123-QED}
\title{Synergistic doping and stabilization of magnetically tunable \textit{Ln}Ti$_3$(Sb,Sn)$_4$ (\textit{Ln}:Ce--Gd) kagome metals}

\author{Brenden R. Ortiz}
\email{ortizbr@ornl.gov}
\affiliation{Materials Science and Technology Division, Oak Ridge National Laboratory, Oak Ridge, TN 37831, USA}

\author{Ramakanta Chapai} 
\affiliation{Materials Science and Technology Division, Oak Ridge National Laboratory, Oak Ridge, TN 37831, USA}
\affiliation{Department of Physics, Norfolk State University, Norfolk, VA 23504, USA}

\author{German Samolyuk} 
\affiliation{Materials Science and Technology Division, Oak Ridge National Laboratory, Oak Ridge, TN 37831, USA}

\author{Milo Sprague} 
\affiliation{Department of Physics, University of Central Florida, Orlando, FL 32816, USA}

\author{Arun K. Kumay} 
\affiliation{Department of Physics, University of Central Florida, Orlando, FL 32816, USA}

\author{Hu Miao} 
\affiliation{Materials Science and Technology Division, Oak Ridge National Laboratory, Oak Ridge, TN 37831, USA}

\author{Karolina G\'{o}rnicka} 
\affiliation{Materials Science and Technology Division, Oak Ridge National Laboratory, Oak Ridge, TN 37831, USA}
\affiliation{Applied Physics and Mathematics Department, Advanced Materials Centre, Gdansk University of Technology, ul. Narutowicza 11/12, 80-233 Gdansk, Poland}

\author{Xiaoping Wang} 
\affiliation{Neutron Scattering Division, Oak Ridge National Laboratory, Oak Ridge, TN 37831, USA}

\author{Qiang Zhang} 
\affiliation{Neutron Scattering Division, Oak Ridge National Laboratory, Oak Ridge, TN 37831, USA}

\author{Madhab Neupane} 
\affiliation{Department of Physics, University of Central Florida, Orlando, FL 32816, USA}

\author{David Parker} 
\affiliation{Materials Science and Technology Division, Oak Ridge National Laboratory, Oak Ridge, TN 37831, USA}

\author{Jiaqiang Yan} 
\affiliation{Materials Science and Technology Division, Oak Ridge National Laboratory, Oak Ridge, TN 37831, USA}

\date{\today}
\begin{abstract} 
Here we present our synthesis and characterization of the \kagSS~(\textit{Ln}: Ce, Pr, Nd, Sm, Gd) family of cleavable kagome metals. While these materials are isostructural to the \textit{Ln}Ti$_3$Bi$_4$ family, they only form as (Sb,Sn) solid-solutions with no corresponding \textit{Ln}Ti$_3$Sb$_4$ or \textit{Ln}Ti$_3$Sn$_4$ phases. We use a combination of first-principles density functional theory (DFT) and Crystal Orbital Hamilton Population (COHP) calculations to show that (Sb,Sn) alloying has a stabilizing effect on the structure by adjusting the Fermi level, filling bonding states, depopulating antibonding states, and adjusting the density-of-states (DOS) towards local minima, an effect we call ``synergistic doping.'' The tunable Fermi level also has a profound effect on the magnetism, which we demonstrate through a detailed characterization of the \SmSS~ series. The series hosts multiple magnetic ground states resulting from competing magnetic interactions that are tunable by the (Sb,Sn) ratio. While the focus of this work is on \SmSS, we briefly comment on the (Sb,Sn) solubility range and the conferred magnetic tunability in the other rare-earths compounds (\textit{Ln}: Ce, Pr, Nd, Gd) as well. Our work demonstrates how the (Sb,Sn) synergistic pair can be used to stabilize the \kagSS~ structure while simultaneously providing a means to tune the magnetism, ultimately providing a potential route to develop new intermetallics with chemical, magnetic, and electronic tunability. 
\end{abstract}
\maketitle
\section{Introduction}

Research into kagome intermetallics continues to represent a key intersection between fundamental condensed matter physics and solid-state chemistry. In particular, the search for materials that host complex magnetic and electronic instabilities has drawn significant attention to the distinctive band structure associated with kagome metals, which is known to generate a unique electronic structure replete with Dirac points, flat bands, and van Hove singularities.~\cite{park2021electronic,PhysRevB.87.115135,kiesel2013unconventional} Naturally, the kagome motif rarely exists in vacuum, and the applicability of these models depends heavily on the actual crystal structure and chemical composition. For example, we often study materials where the ``contaminating'' effect of other orbitals and supporting sublattices can be minimized, often by controlling the dimensionality of the compound.\cite{jovanovic2022simple} Another crucial parameter that parametrizes a kagome compound is the alignment of the Fermi level with key features in the electronic structure. Thus, there are a wide array of chemical and structural considerations which must be made to enable the latent potential of the kagome motif in real materials.

Invigorated in part by our earlier discovery of the \textit{A}V$_3$Sb$_5$ (\textit{A}: K, Rb, Cs) kagome metals \cite{ortiz2019new,ortizCsV3Sb5}, a huge body of research has emerged investigating emergent topological, electronic, and magnetic instabilities (e.g. charge-density waves and superconductivity) within kagome metals.\cite{wilson2024v3sb5,wang2023quantum,yin2022topological,neupert2022charge,di2026kagome,jiang2023kagome,wang2024topological} Unfortunately, it seems that the chemical flexibility of the \textit{AM}$_3$\textit{X}$_5$ family is limited. Since the original discovery, only three other materials have been discovered: CsCr$_3$Sb$_5$\cite{liu2023superconductivity}, CsTi$_3$Bi$_5$\cite{werhahn2022kagome}, and RbTi$_3$Bi$_5$\cite{werhahn2022kagome}. This scarcity has helped ignite a wide range of exploratory studies searching for new kagome intermetallics exhibiting density-wave instabilities. Materials based in the extended CoSn-family (including \textit{AM}$_6$\textit{X}$_6$ materials) are the most topical examples, with recent discoveries of density-wave behavior exhibited in LuNb$_6$Sn$_6$,\cite{ortiz2025stability} ScV$_6$Sn$_6$,\cite{arachchige2022charge} and FeGe.\cite{teng2022discovery} However, unlike the more electronically-driven behavior in the \textit{AM}$_3$\textit{X}$_5$ family, the CoSn-based derivatives appear to be driven by a combination of steric effects that affects only a small subset of the known compounds.\cite{Hu2024_PhononPromotedCDW-ScV6Sn6,Meier2023tiny,Pokharel2023_FrustratedCO+CooperativeDistortionsScV6Sn6,Cao2023_CompetingChargeOrderScV6Sn6,Lee2024_NatureCDWScV6Sn6,Korshunov2023_SofteningPhononScV6Sn6,Hu2023_ScV6Sn6-Theory-FlatPhonons+UnconventionalCDW,Liu2024_DrivingMechanismScV6Sn6,Yu2024_MagAndCorrelationsScV6Sn6,Wang2023_EnhancedSpinPolarizationViaDimerizationFeGe,Wen2024_UnconventionalCDW-FeGe,Chen2024_LongRangeGeDimerizationFeGe} 

Another rapidly developing family of cleavable kagome metals is the \textit{AM}$_3$\textit{X}$_4$ family. The \textit{AM}$_3$\textit{X}$_4$ materials contain two key structural motifs of interest: 1) zig-zag chains of \textit{A} atoms, and 2) kagome nets formed by $M$ ions. The \textit{A} site is often occupied by electropositive rare-earth or alkali-earth metals, enabling a wide array of highly anisotropic, low-dimensional magnetic properties. Combined with the electronic structure imparted by the kagome network, these materials have a high potential for magnetic, electronic, and structural interactions. We previously reported on the discovery of the antimonide\cite{ortiz2023ybv} \textit{Ln}V$_3$Sb$_4$ (\textit{Ln}: Eu$^{2+}$, Yb$^{2+}$) and bismuthide\cite{ortiz2024intricate, ortiz2023evolution} \textit{Ln}Ti$_3$Bi$_4$ (\textit{Ln}: La...Tb$^{3+}$, Eu$^{2+}$, Yb$^{2+}$) kagome metals. These materials are part of a larger class of \textit{AM}$_3$\textit{X}$_4$ kagome metals reported in recent years.\cite{ortiz2023ybv,ortiz2019new,ortiz2023evolution,ovchinnikov2018synthesis,ovchinnikov2019bismuth,motoyama2018magnetic,chen2023134,guo2023134}

Even more recently, reports have emerged suggesting the emergence of spin- and charge-density anomalies in TbTi$_3$Bi$_4$,\cite{TbCDW_cheng2024spectroscopic} GdTi$_3$Bi$_4$,\cite{GdCDW_han2025discovery} and CeTi$_3$Bi$_4$.\cite{CeCDW_park2025spin} In the case of CeTi$_3$Bi$_4$, there is good evidence for a van Hove singularity-assisted spin-density wave, demonstrating the intertwining of magnetic and electronic interactions. Particularly within the bismides, the availability of large, cleavable single crystals has made the \textit{AM}$_3$\textit{X}$_4$ systems amenable to many experimental techniques. A number of angular-resolved photoemission (ARPES),\cite{NdARP_mondal2025observation,NdARP_hu2024magnetic,SmARP_zheng2024anisotropic,EuARP_jiang2024topological,GdARP_cheng2024striped,TbARP_zhang2024observation,TbARP_kushnirenko2025observation,YbARP_sakhya2024diverse,LaARP_sakhya2025diverse,TbSTM_zhang2025observation} scanning-tunneling microscopy (STM),\cite{TbSTM_zhang2025observation,GdCDW_han2025discovery} and other complementary techniques (e.g. magnetic force microscopy, torque magnetometry, and neutron diffraction)\cite{GdMFM_guo2025tunable,TbSTM_zhang2025observation,YbOsc_shtefiienko2025electronic} have investigated the connection between the electronic structure and the rare-earth magnetism. Concomitant with the complex magnetic properties, a wide array of complex magnetotransport phenomena\cite{TbCDW_cheng2024spectroscopic,GdXport_li2025anisotropic,EuXport_shu2025complex,ortiz2024intricate} have been observed, including a record-large anomalous Hall effect (AHE) in TbTi$_3$Bi$_4$.\cite{TbCDW_cheng2024spectroscopic} Particularly given the relative youth of the \textit{AM}$_3$\textit{X}$_4$ family, these materials appear to be a fertile ground for exploring the interaction between magnetism and the underlying kagome band structure.

This work focuses on the synthesis and characterization of an analogous series of materials, the \kagSS~ (\textit{Ln}: Ce--Gd) metals. These materials exist only as a solid-solution between the hypothetical \textit{Ln}Ti$_3$Sn$_4$ and \textit{Ln}Ti$_3$Sb$_4$. Accordingly, we first explore the stabilizing effect of the (Sb,Sn) alloying using a combination of density-functional theory (DFT) and crystal orbital Hamilton population (COHP) calculations. We demonstrate that the (Sb,Sn) pair electronically stabilizes the structure by controlling electron filling while not dramatically altering the electronic structure. We dub this effect ``synergistic doping'' and subsequently propose a synthetic strategy based on synergistic pairs. The variable Fermi level also manifests as tunable magnetism, which we demonstrate through a detailed investigation of the \SmSS~ physical properties. We show that the system exhibits both ferromagnetic (FM) and antiferromagnetic (AFM) signatures, and that the (Sb,Sn) ratio can control the blending and dominance of the interactions. We also briefly comment on the (Sb,Sn) control and analogous properties in the other rare-earth (Ce, Pr, Nd, Gd) compounds as well. Our results ultimately show how the (Sb,Sn) synergistic dopants can stabilize the \kagSS~ crystal structure, control the magnetic ground state, and provide a potentially new synthetic strategy for the discovery of complex intermetallic compounds.

\section{Methods}
\subsection{Synthesis}
\kagSS (\textit{Ln}: Ce--Gd) single crystals are grown from alloys of (Sb,Sn) using the self-flux method. Rare-earth metal pieces (Ames) were placed with Ti powder (Alfa 99.9\%), Sb shot (Alfa 99.999\%), and Sn shot (Alfa 99.9\%) into 2~mL Canfield crucibles fitted with a catch crucible and a porous frit.\cite{canfield2016use}. The ratio of \textit{Ln} and Ti was fixed relative to the total amount of (Sb,Sn) flux such that the total composition adhered to \textit{Ln}$_{0.25}$Ti$_{0.75}$Sb$_{x}$Sn$_{10-x}$. The specific (Sb,Sn) flux compositions that yield single crystals is discussed later.

The filled crucibles were sealed under approximately 0.7~atm of argon gas in fused silica ampoules. Samples were heated to 1050\degree C at a rate of 200\degree C/hr and thermalized at 1050\degree C for 12~h. The vast majority of samples are cooled to 750\degree C at a rate of 2\degree C/hr. However, we observe that the most Sn-rich compositions often require further cooling to 650\degree C to form the correct phase. Excess (Sb,Sn) flux was removed through centrifugation at either 750\degree C or 650\degree C. 

Single crystals of \kagSS~ are silver-colored hexagonal plates with a metallic luster. Sample sizes commonly range from 1-10~mm depending on the rare-earth species and the specific ratio of (Sb,Sn). Unlike the structurally analogous \LTB~ compounds, \kagSS~ are completely stable in air and atmospheric moisture. Samples have remained untarnished on the scale of 6~months, unlike the \LTB~ compounds which will tarnish and spall within a few hours in air. \kagSS~ crystals retain their easy (00L) cleavage plane, and can be readily cleaved with a sharp blade. Cleaved surfaces can be further exfoliated using adhesive tape.

\subsection{Bulk Characterization}
Elemental analysis of single crystals was carried out on as-grown crystals using a Hitachi-TM3000 microscope equipped with a Bruker Quantax 70 EDS system. Crystals were mounted on carbon tape and cleaved to reveal pristine, flat surfaces before elemental analysis was performed. Subsequently, single crystals of \kagSS~ were mounted on Kapton loops with Paratone oil for single crystal x-ray diffraction (SCXRD). Diffraction data were collected at 300~K on a Bruker D8 Advance Quest diffractometer with a graphite monochromator using Mo K$\alpha$ radiation ($\lambda$ = 0.71073~\AA). Data integration, reduction, and structure solution was performed using the Bruker APEX4 software package. As noted in previous manuscripts,\cite{ovchinnikov2019bismuth,ortiz2023evolution} SCXRD is a challenging endeavor on these materials due to their naturally large sizes, extremely soft mechanical properties, and exfoliable nature. Due to the difficulty of resolving Sb and Sn in X-ray diffraction, we also performed preliminary neutron diffraction measurements on Sb-rich crystals of NdTi$_3$(Sb,Sn)$_4$ (295~K) on the TOPAZ beamline at the Spallation Neutron Source, Oak Ridge National Laboratory.\cite{coates2018suite} Within resolution of our measurement, we were not able to find significant evidence for  preferential occupancy of (Sb,Sn), and no structural superlattices were observed. 

Magnetization measurements (300 -- 1.8~K) on crystals of \kagSS~ were performed in a 7~T Quantum Design Magnetic Property Measurement System (MPMS3) SQUID magnetometer in vibrating-sample magnetometry (VSM) mode. Heat capacity measurements (300--1.8~K) were performed in a Quantum Design 9~T Dynacool Physical Property Measurement System (PPMS). Heat capacity measurements utilized the large-pulse (30\% temperature rise) method, subsequently processed by both the dual-slope and single-slope method. Electronic transport measurements were performed in a Quantum Design 12~T Dynacool PPMS equipped with the Electronic Transport Option (ETO). An alternating current of 10\,mA and 100\,Hz frequency was applied as the probe current for magnetoresistance measurements. 

\subsection{Electronic Structure}
Angle-resolved photoemission spectroscopy (ARPES) measurements were carried out at the Advanced Light Source (ALS) at Lawrence Berkeley National Laboratory. Data were collected at beamline 10.0.1.1 using a Scienta R4000 hemispherical electron analyzer with an incident photon energy of 60~eV. The angular and total energy resolutions were set to 0.2° and 15 meV, respectively. High-quality single crystals were cut and mounted on copper posts using silver epoxy, with ceramic posts affixed to the samples. After loading into the main chamber, the samples were cooled and maintained under vacuum for several hours, before cleaving the sample by striking the ceramic post. Sample cleavage and all subsequent measurements were performed at a sample temperature of 20~K under ultra-high vacuum conditions ($<$10$^{-10}$~Torr). Supplementary ARPES was performed at beamline 21-ID-1 of NSLS-II at Brookhaven National Lab (BNL) with an incident energy of 100~eV, linearly polarized photons, and 15~meV energy resolution.

DFT results utilized self-consistent methods\cite{kohn1965self} to evaluate the electronic structure of \SmSS. The electronic exchange correlations were incorporated using generalized gradient approximation (GGA) within the Perdew-Burke-Ernzerhof (PBE) \cite{perdew1996} parametrization.  The plane-wave basis projector augmented-wave approach (PAW) \cite{blochl1994projector} as implemented in the Vienna Ab-initio Simulation Package (VASP) \cite{kresse1996efficiency,kresse1999ultrasoft} was used. We used a plane-wave energy cutoff of 450~eV and a 6$\times$6$\times$6 k-point mesh to execute the self-consistent calculations. The Sm$^{3+}$ pseudopotential describes electron ion interaction for Sm atoms and is accurate enough to describe details of the band structure in the nonmagnetic state. For Ti, the pseudopotential used treats the 3p semicore states as valence states. The original structure atomic positions were optimized until the energy change was below 0.05 meV and forces were below 0.5 meV/$\AA$. 

For a chemical interpretation of the bonds within \SmSS, we employed the Crystal Orbital Hamilton Population (COHP) analysis. This technique allows us to represent the decomposition of the band structure over pair projected orbitals\cite{COHP_deringer2011crystal,COHP_dronskowski1993crystal} and ultimately allows us to identify bonding and antibonding contributions pairwise atomic interactions. The COHP calculations utilize our VASP results\cite{maintz2013analytic} as inputs to the LOBSTER COHP package.\cite{maintz2016lobster} Postprocessing of the LOBSTER results was performed using LOPOSTER.\cite{wang2025loposter}

\section{Results and Discussion}
\subsection{Synthesis and Crystal Structure}

Initial attempts to prepare the \textit{Ln}Ti$_3$Sb$_4$ analogs of the \textit{Ln}Ti$_3$Bi$_4$ phases\cite{ortiz2019new,ortiz2023evolution,ovchinnikov2018synthesis,ovchinnikov2019bismuth,motoyama2018magnetic,chen2023134,guo2023134} \textit{via} Sb self-flux were unsuccessful. However, using a Sn-based flux produced the structurally analogous \kagSS~(\textit{Ln: Ce--Gd}) series of alloys. An abbreviated representation of the crystal structure is shown in Figure \ref{fig:xtal}(a) and CIF files for a representative selection of the \kagSS~ compounds is provided in the supplemental information. The structure has two sublattices of interest:  1) the \textit{Ln}-based zig-zag chains, and 2) the Ti-based kagome network, both highlighted in Figure \ref{fig:xtal}(b). Compared to the corresponding bismuthide, the unit cell shrinks approximately 7\% (SmTi$_3$Bi$_4$:1500$\text{\AA}^3$, Sb-rich \SmSS: 1400$\text{\AA}^3$), which is reflected in the associated Ti--Ti and \textit{Ln}--\textit{Ln} interatomic distances. Owing to the stronger Sb--Ti bonds, the \kagSS~ appear to be stable in air and atmospheric water for up to a year without tarnishing, a dramatic improvement on the air-sensitive ($<$1~hr.) bismuthides.

As remarked before, the \kagSS~ series appears to contain a substantial amount of (Sb,Sn) mixing, and we were unable to isolate either the pure antimonide or the pure stannide using flux methods. A prior report from Bie \textit{et. al} also notes their inability to form either SmTi$_3$Sb$_4$ or NdTi$_3$Sb$_4$ using solid-state methods,\cite{bie2007ternary} suggesting that these phases may not exist. Using (Sb,Sn) alloys as the flux, we prepared single crystals \kagSS~ for the extended series from \textit{Ln}:(Ce--Gd), and further identified the solubility range under a variety of (Sb,Sn) ratios and synthetic conditions.

Figure \ref{fig:xtal}(c) summarizes the range of accessible (Sb,Sn) alloys for each of the rare-earth elements investigated here. The Sb-rich side of the diagram highlights that the compounds are nearly pure antimonides. Figure \ref{fig:xtal}(d), which shows the nominal (Sb,Sn) flux composition against the actual (Sb,Sn) incorporation in the single crystals, further highlights the \kagSS~ favoritism towards Sb-rich compositions. Even at 10\% nominal Sb loading in the flux, the (Sb,Sn) site is 80\% Sb, though there are diminishing returns as more Sb is added to the flux. After 40-50\% doping, even the Sm-containing series terminates with the appearance of other ternary and quaternary phases. We hypothesize that the reduced (Sb,Sn) solubility in the Ce and Pr compounds may be a consequence of a competing ternary antimonide \textit{Ln}$_2$Ti$_9$Sb$_{11}$ (\textit{Ln}: La--Nd) we reported  previously.\cite{ortiz2025isolated} This may also explain the absence of a La-containing \kagSS, as we previously showed that the competing \textit{Ln}$_2$Ti$_9$Sb$_{11}$ is stabilized most by the larger La ion. 

\begin{figure}
\includegraphics[width=\linewidth]{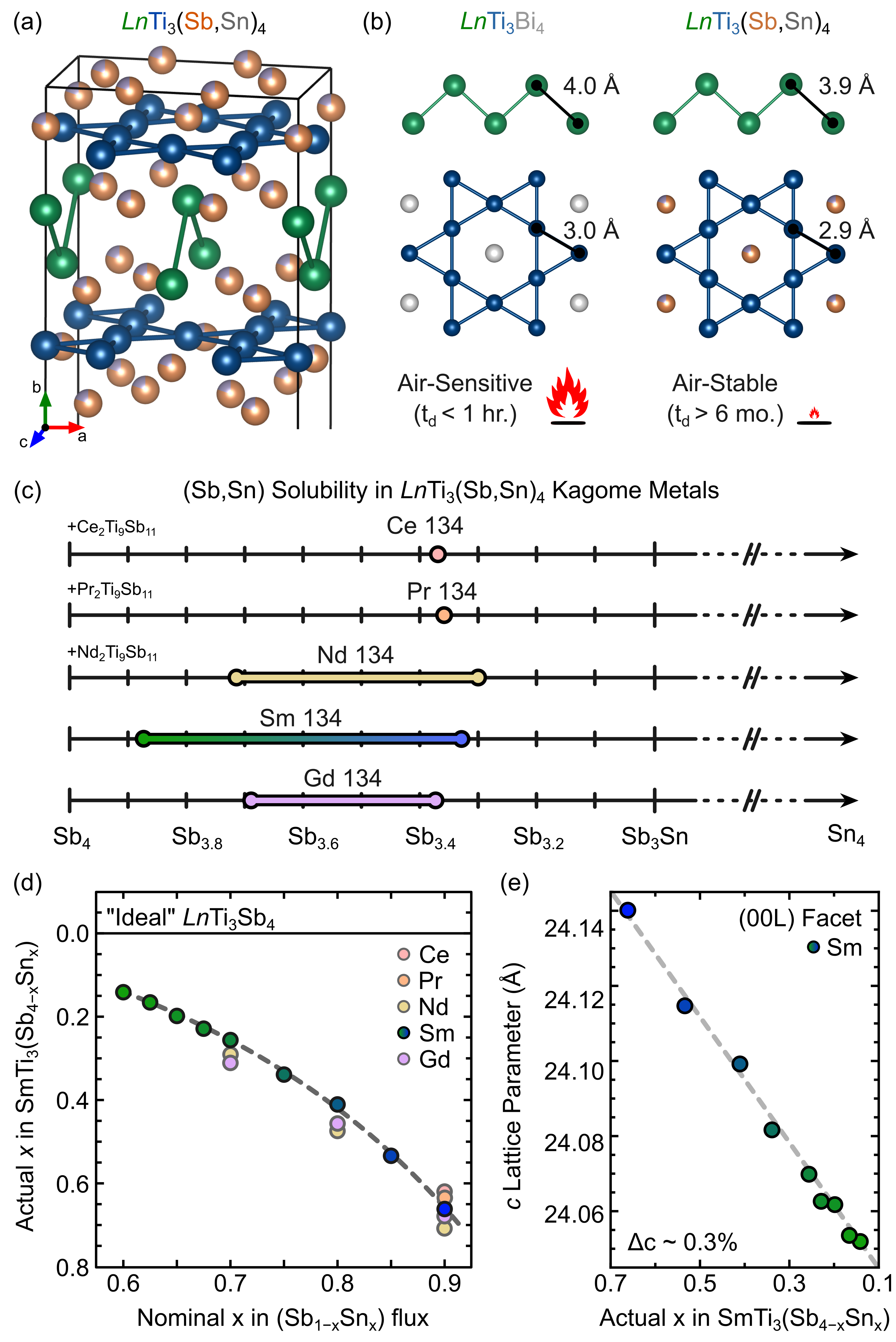}
\caption{(a) Abbreviated view of the \textit{Fmmm} \kagSS~ crystal structure. (b) A schematic comparison of the structural motifs in the \textit{Ln}Ti$_3$Bi$_4$ and \kagSS~ families. (c) (Sb,Sn) solubility ranges for each of the flux-grown \kagSS~ single crystals. We've also highlighted compositions where competition with \textit{Ln}$_2$Ti$_9$Sb$_{11}$ phase is observed. (d) Processing diagram showing the incorporation of (Sb,Sn) into the crystal structure depending on the nominal Sb percentage in the flux. (e) Shift in the \SmSS~ \textit{c}-axis lattice parameter as a function of composition, derived from \{00L\} facet scans.}
\label{fig:xtal}
\end{figure}

\begin{figure*}
\includegraphics[width=\textwidth]{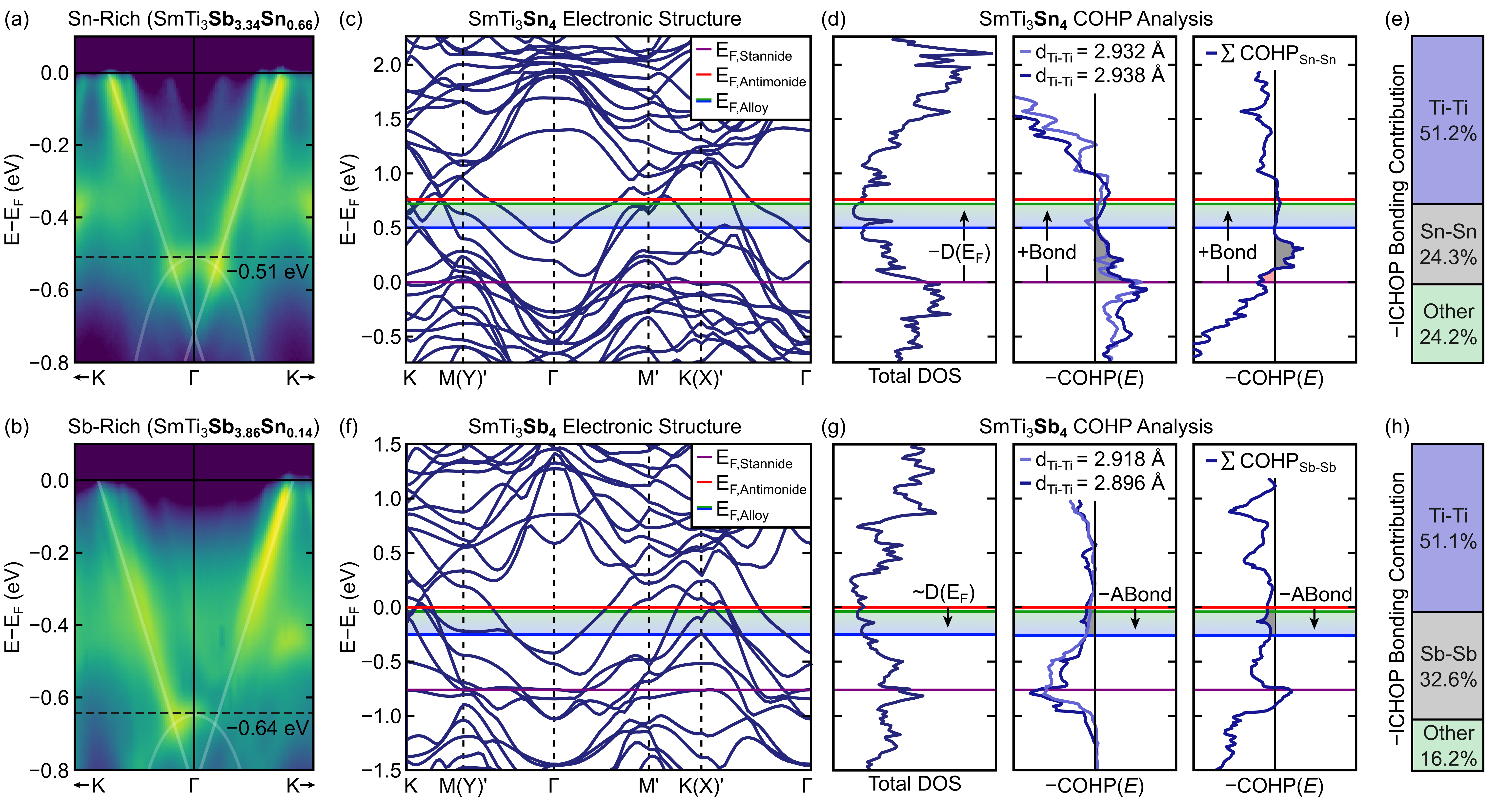}
\caption{(a) Experimental ARPES data for a Sn-rich \SmSS~ crystal, with several bands highlighted in white. Note the concave-down band approximately 0.5~eV below E$_\text{F}$. (b) Analogous ARPES data for a Sb-rich \SmSS~ crystal. The concave-down band is now 0.64~eV below E$_\text{F}$, indicating an effective doping of approximately 270~meV per Sb/Sn atom. (c) Electronic band structure for the hypothetical SmTi$_3$Sn$_4$. Several values of \ef~ are shown: the native stannide (purple), the native antimonide (red), and the range approximating the alloying range (green/blue). (d) Density of states (D(E)) and Crystal Orbital Hamilton Population (COHP) calculations for SmTi$_3$Sn$_4$ showing that electron doping (Sb-substitution) in the stannide is favorable through reduction of D(\ef) and population of additional bonding states. (e) Breakdown of integrated (ICOHP) contribution to the total bonding band energy for Ti-Ti and Sn-Sn bonds (shown) relative to all other atom pairs. (f) Dispersion for hypothetical SmTi$_3$Sb$_4$ shown with same set of \ef~ as before. (g) Analogous D(E) and COHP calculations for the antimonide. Hole doping (Sn-substitution) in antimonide is favorable through reduction of antibonding states at E$_\text{F}$ while maintaining low D(\ef). (h) Breakdown of integrated ICOHP contribution for antimonide.}
\label{fig:DFT}
\end{figure*}

In standard lab X-ray diffraction (XRD), it is very difficult to distinguish Sb and Sn due to their negligible Z-contrast. To check the distribution of (Sb,Sn) we performed a preliminary neutron diffraction experiment on crystals of \NdSS~ and found no indications of a preferential distribution of (Sb,Sn). As an additional check against our energy dispersive spectroscopy (EDS) results, we performed (00L) facet scans on cleaved crystals of \SmSS~ which provides a very accurate determination of the \textit{c}-axis lattice parameter. Despite the similar size and chemistry of (Sb,Sn), we still expect a weak contraction of the unit cell as the Sb concentration increases. Figure \ref{fig:xtal}(e) confirms the very weak (0.3\% change) in the \textit{c}-axis lattice parameter across the entire \SmSS~ alloy series. For comparison, this effect is over an order-of-magnitude weaker than the volumetric change across the rare-earth series (3.5\% between \CeSS~ and \GdSS~). The corresponding plot of the cell volume versus the 9-coordinate Shannon radius can be found in the supplementary information. 

We see no evidence of peak splitting in facet scans and no compositional segregation in energy dispersive spectroscopy (EDS), indicating good structural homogeneity throughout the 1-5~mm sized crystals. Crystals within a single batch exhibit consistent stoichiometry and properties, and general batch-to-batch behavior can be well-reproduced. We have observed that altering the ratio of the starting reagents (e.g. \textit{Ln}$_{0.25}$Ti$_{0.75}$Sb$_{x}$Sn$_{10-x}$ vs \textit{Ln}$_{0.25}$Ti$_{0.75}$Sb$_{2x}$Sn$_{20-2x}$) will change the correspondence of actual:nominal, but will not enable further incorporation of Sb in the \kagSS~ series. From the uniform expansion of the unit cell, the lack of peak splitting, the uniform spectroscopy results, and our synthetic observations, we believe that the crystals are macroscopically homogeneous alloys. There is always the potential for short-range (Sb,Sn) fluctuations, however, and more detailed measurement techniques (e.g. MFM or STM) could be interesting routes to investigate the local effect of the (Sb,Sn) doping on the electronic and magnetic properties.

For the remainder of this manuscript we will focus on the \SmSS~ series. We choose \SmSS~ because it shows the largest variation in (Sb,Sn) incorporation, exhibits a strong and complex magnetic ground state with a relatively high transition temperature (10-20~K), and exhibits anisotropy (easy \textit{c}-axis) which is easily oriented for. Throughout the paper we will refer to the Sb-rich terminus of the Sb-series (SmTi$_3$Sb$_{3.86}$Sn$_{0.14}$) simply as Sb-rich \SmSS~, and it will be colored green in all subsequent plots. Similarly, the Sn-rich composition (SmTi$_3$Sb$_{3.34}$Sn$_{0.66}$) will be referred to as Sn-rich \SmSS~ and will be colored blue. A brief discussion of the other rare-earth elements \kagSS~ \textit{Ln}:(Ce,Pr,Nd,Gd) is included at the end of the manuscript with basic characterization data included in the supplementary information.

\subsection{``Synergistic Doping'' and Stability of \kagSS}

Both our work and work from Bie et. al\cite{bie2007ternary} suggest that the \kagSS~ compounds exist only as an alloy between the hypothetical compounds \textit{Ln}Ti$_3$Sb$_4$ and \textit{Ln}Ti$_3$Sn$_4$. We hypothesize that the structure may be stabilized by a certain electron count, and that the chemical (radii, electronegativity, bonding) similarity of Sb and Sn allows the system to naturally equilibrate at the electron count that stabilizes the structure. For brevity, we are calling this effect -- imbuing a system with chemically similar alloying agents to enable a natural charge degree-of-freedom and stabilize a structure -- ``synergistic doping.'' In this section we will combine experimental angular-resolved photoemission spectroscopy (ARPES) and first-principles density functional theory (DFT) to examine whether a potential charge degree-of-freedom may explain the stabilization of the \kagSS~ series. As mentioned before, we will focus on the Sm-containing series for the remainder of the discussion. 

First, we establish that (Sb,Sn) enables tunability of the Fermi level (\ef) in \SmSS~ without drastically changing the electronic structure. To benchmark changes in \ef~ induced by the alloying, we performed experimental ARPES measurements on the Sn-rich (SmTi$_{3}$Sb$_{3.34}$Sn$_{0.66}$) and Sb-rich (SmTi$_{3}$Sb$_{3.86}$Sn$_{0.14}$) single crystals. These results are shown in Figure \ref{fig:DFT}(a) and Figure \ref{fig:DFT}(b), respectively. To determine the doping shift of \ef, we examine the intersection of a series of dispersive bands crossing through $\Gamma$ with another concave-down band.  The bands have been highlighted in the figure with a light gray trace and the intersection has been marked with a dashed black line. Subtracting the intersection energy in the Sn-rich data (-0.51~eV) from the Sb-rich data (-0.64~eV) yields a shift of 130~meV. Normalizing this shift to the (per atom) concentration difference between SmTi$_3$Sb$_{3.34}$Sn$_{0.66}$ and SmTi$_3$Sb$_{3.86}$Sn$_{0.14}$, we observe a shift of 260~meV per (Sb,Sn) substitution.

A similar experiment can be performed with DFT, where we simulate the expected shift in \ef~ under the effect of (Sb,Sn) doping. We will investigate the effect of Sb-doping on SmTi$_3$Sn$_4$ and Sn-doping on SmTi$_3$Sb$_4$. While these compounds do not formally exist, this alleviates the complexity of alloy simulations in low-symmetry, large unit cell systems and further lets us examine the differences between the electronic structures. Figure \ref{fig:DFT}(c,f) show the DFT calculations for SmTi$_3$Sn$_4$ and SmTi$_3$Sb$_4$ over the pseudo-hexagonal Brillouin zone (for details on the pseudo-hexagonal zone projection, see our discussions in Ref \cite{ortiz2023evolution,CeCDW_park2025spin}). For each simulation we show three key values of\ef: (1) the native \ef~ for the pure Stannide (purple), (2) the native \ef~ for the pure antimonide (red), and (3) a range of potential \ef~ given the (Sb,Sn) alloying (blue/green). To determine the \ef~ window for the alloys, we used the compositions SmTi$_3$Sb$_{3.1}$Sn$_{0.9}$ and Sb-rich SmTi$_3$Sb$_{3.9}$Sn$_{0.1}$ and assumed that each (Sb,Sn) substitution results in 1 electron/hole per swap. This restraint yields a computed \ef~ window of 210~meV (a per atom basis of 263~meV per swap) which is in excellent agreement with experimental ARPES data. 

The DFT results in Figure \ref{fig:DFT}(c,f) also allow us to examine the general influence of (Sb,Sn) alloying on the bulk band structure. Qualitatively, the band structures of SmTi$_3$Sn$_4$ and SmTi$_3$Sb$_4$ are remarkably similar near the relevant \ef, despite full substitution of Sb and Sn. This suggests that the (Sb,Sn) swapping likely has a minimal impact on the local electronic structure, particularly near \ef. This is perhaps expected, since many kagome materials tend to have the states at \ef~ dominated by the \textit{M}$_3$ atoms. In the supplemental information we have included a series of additional ARPES measurements that show further evidence for the innocuous nature of the (Sb,Sn) substitution. In short, we've observed that constant-energy-contours (CEC's) generated around -130~meV in the Sb-rich compound (corresponding to the proposed doping level) produce CEC's which are very similar to the CEC for the Sn-rich compound at it's natural \ef. From these results, we believe that the alloying does not impact the bulk electronic structure in any substantial manner and primarily acts to tune \ef.

We're now prepared to tackle \textit{why} the \kagSS~ compounds are stabilized by Fermi level tuning. For this discussion we will combine density of states (D(E)) calculations and Crystal Orbital Hamilton Population (COHP) to provide a qualitative argument for the enhanced stability of the alloys. Prefacing our discussion, we intuitively expect that materials \textit{generally} crystallize such that \ef~ is placed near a local minimum in D(E). While not a strict rule, placing \ef~ in such a way tends to reduce the system's total energy. Systems with \ef~ at local maxima tend to distort or destabilize in a way that restructures the DOS and avoids high D(\ef). However, D(E) only provides the number of states and not the chemical character (e.g. bonding or antibonding) of the orbitals. For this we turn to COHP simulations, which allow us to identify specific bonding, antibonding, and nonbonding contributions to the electronic structure. In the typical presentation, COHP is presented as (-COHP(E)) where positive values correspond to bonding states. Note that our discussion doesn't include any arguments from configurational entropy, although those would generally be believed to enhance the energetic stability of an alloy as well.

Figure \ref{fig:DFT}(d) presents D(E) and COHP(E) calculations for SmTi$_3$Sn$_4$. For clarity our discussion here will focus on the Ti--Ti and Sn--Sn interactions. These are two largest contributors to the total off-site bonding interactions (Figure \ref{fig:COHP}(e,h) provide the relative contributions to the COHP), though all interactions are provided in the supplementary information. There are two unique Ti--Ti bond distances shown in Figure \ref{fig:COHP}(d), though we have summed the nine individual Sn--Sn interactions for graphical clarity. From D(E), we can see that the native Fermi level (purple) is near a local maximum in D(E). Electron doping \textit{via} Sb substitution (purple to green/blue) adjusts \ef~ up and into a local minimum. Furthermore, electron doping fills additional bonding states (shaded gray) in both the Ti--Ti and Sn--Sn COHP calculations, which further stabilizes the structure. The system reaches a natural limit in the electron-doped regime as further doping would increases D(\ef) and incorporate substantial antibonding contributions. 

An analogous effect can be observed in the antimonide. However, since the alloy range is very close to the antimonide limit, the effect is appreciably more subtle. Figure \ref{fig:DFT}(g) examines the DOS and COHP simulations for SmTi$_3$Sb$_4$ near the relevant energies. Unlike the stannide, the antimonide already sits near a relatively flat local minimum in D(E). Examining the dominant Ti--Ti and Sb--Sb COHP contributions reveals that the natural Fermi level in the antimonide naturally falls in a heavily antibonding regime. The antimonide feels two competing effects: the desire to keep D(E) low, and the desire to depopulate antibonding states by pushing \ef~ down. The interplay between these two effects culminates as justification for the alloying range (green/blue) sitting slightly below the antimonide's natural \ef~ (red). The system likely naturally self-limits the Sn doping due to increasingly large D(\ef) under hole doping, even though the driving force to continue depopulating the antibonding states remains. 

From our analysis, it seems plausible that the charge degree-of-freedom offered by the (Sb,Sn) pair may justify the formation of the \kagSS~ alloying series and the apparent difficulty in synthesizing the pure antimonide. We've demonstrated that (Sb,Sn) alloys substitute with nearly 1:1 doping efficiency, shifting the experimental Fermi level in accordance with simulations while leaving the bulk band structure largely unchanged near \ef. This combination of effects imbues the system with the charge flexibility needed to adjust the Fermi level, balancing energetic considerations from both the DOS and the bonding/antibonding COHP analysis. Employing the strategy of ``synergistic doping'' may have considerable implications on the discovery and development of other structurally complex intermetallics.

\begin{figure*}
\includegraphics[width=1\textwidth]{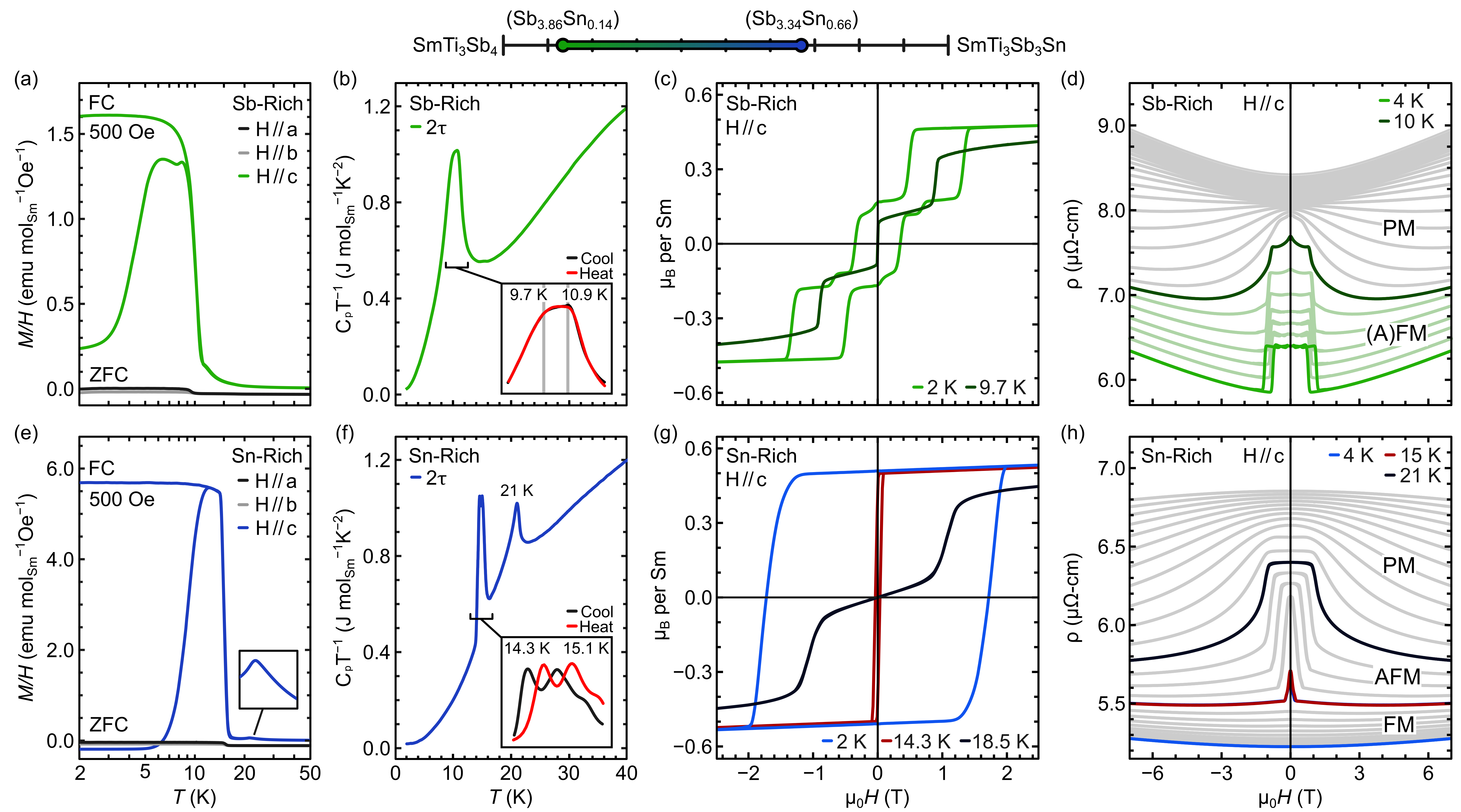}
\caption{Here we present the basic properties of the Sb-rich (green) and Sn-rich (blue) termini of the \SmSS~ solid-solution. (a,e) Field-cooled magnetization along three principle crystallographic directions, showing easy \textit{c}-axis magnetization. Both systems show a higher-temperature cusp followed by a lower-temperature magnetization rise and plateau. (b,f) Large-pulse zero-field heat capacity measurements showing the multiple transitions in both compositions. Sn-rich samples a second order peak at 21~K followed by a split first-order peak at 15~K. Sb-rich samples instead show a set of two second-order peaks at 9.7~K and 10.9~K.
(c,g) Isothermal magnetization results with H$\parallel$c clearly demonstrate the FM nature of the Sn-rich sample and the more complex A(FM) interactions in the Sb-rich endpoint. (d,h) Magnetoresistance measurements showing the emergence of negative magnetoresistance (MR) in the AFM state, which persists through 4~K in the Sb-rich sample. The onset of stronger FM order in the Sn-rich sample suppresses the AFM and the negative MR, consistent with the first-order nature of the AFM to FM transition Sn-rich heat capacity.}
\label{fig:SmBasic}
\end{figure*}

\subsection{Properties of \SmSS~ solid solution}

\begin{figure}
\includegraphics[width=\linewidth]{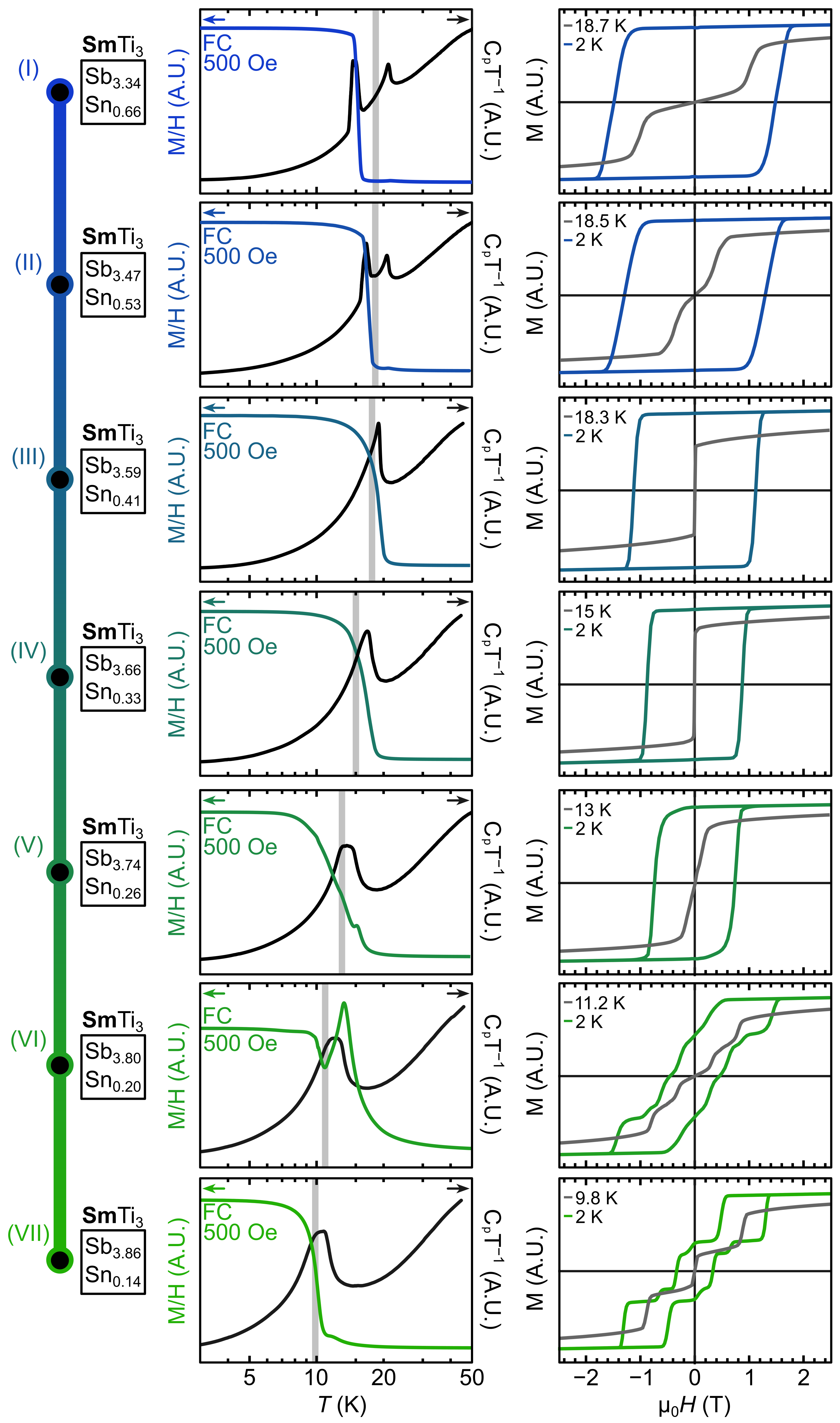}
\caption{Here we present a full series of samples traversing the \SmSS~ solid-solution to demonstrate the gradual shift in magnetic and thermodynamic properties. Shown from left to right are: (1) the measured (Sb,Sn) content of the single crystals, (2) a plot showing the temperature-dependent magnetization and zero-field heat capacity results superimposed and scaled in arbitrary units, and (3) the corresponding isothermal magnetization curves at base temperature and an intermediate temperature (gray) intended to demarcate the transition between the two heat capacity features (where possible). For clarity, the intermediate temperature (gray) is also marked on the temperature-dependent magnetization.}
\label{fig:SmGrid}
\end{figure}

\begin{figure*}
\includegraphics[width=1\textwidth]{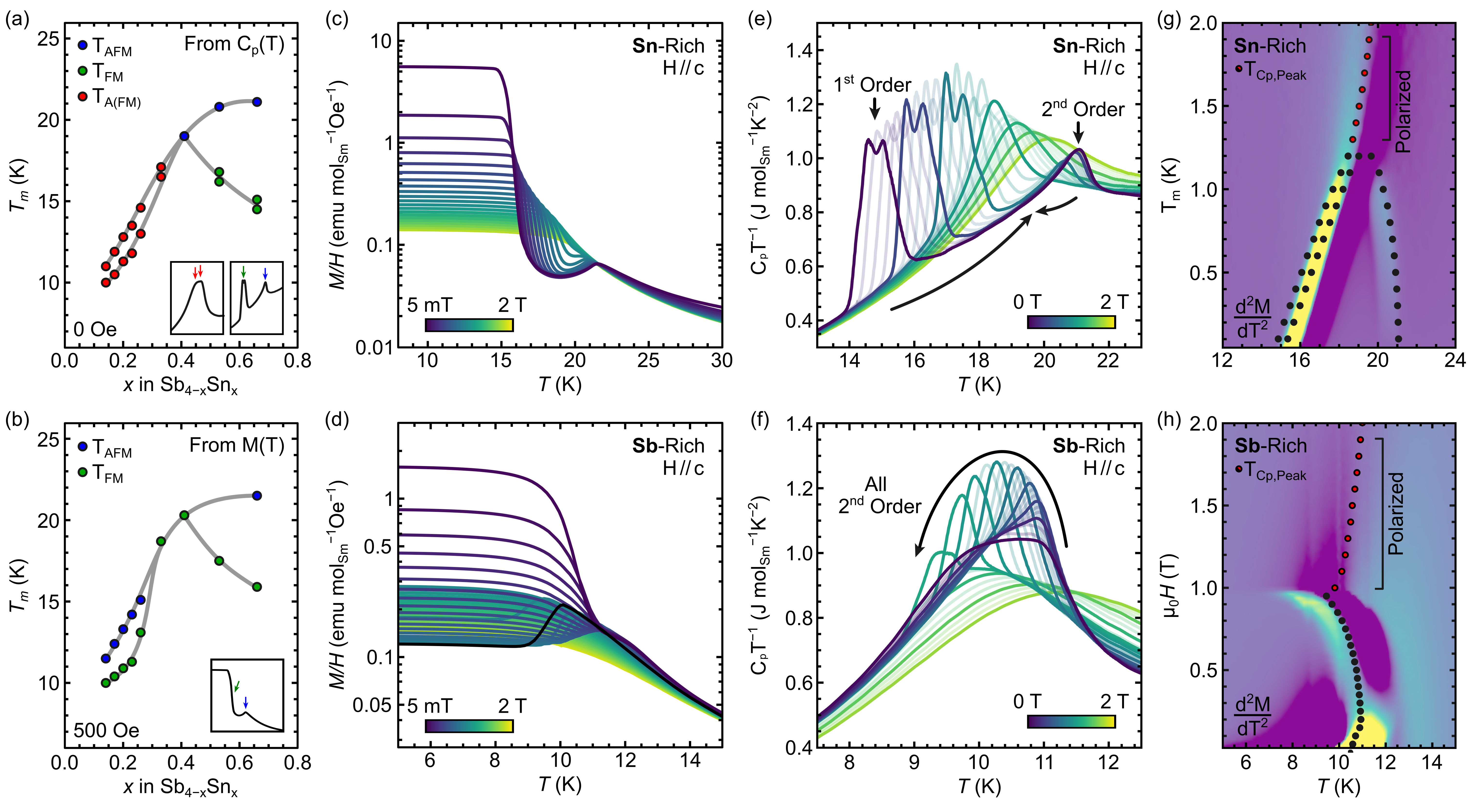}
\caption{Here we summarize the temperature-composition phase diagrams for the \SmSS~ solid-solution, and further investigate the field-temperature phase diagrams for the Sb-rich and Sn-rich termini of the alloys. Temperature-composition phase diagrams derived from (a) heat capacity and (b) magnetization measurements demonstrate the competition between the higher-temperature AFM order and the onset of FM order at lower temperatures. For Sb-rich compositions, the AFM order and FM interactions intertwine into the A(FM) state. (c,d) Temperature-dependent magnetization curves for the Sn-rich and Sb-rich termini at under various magnetic fields. The AFM-like cusp in the Sn-rich sample is evident and appears gradually suppressed under applied field. The AFM-like cusp in the Sb-rich sample is more subtle at low fields, but is strengthened and more dominant at intermediate fields (highlighted in black for clarity). (e,f) Field-dependent large-pulse heat capacity measurements to help visualize the multiple heat capacity peaks and their interplay with fields. (g,h) Composite plots where the gradient background is the second-derivative of the field- and temperature-dependent magnetization. Solid points are identified from the field- and temperature-dependent heat capacity results.}
\label{fig:SmPhaseDia}
\end{figure*}

The potential for kagome materials to host a suite of complex magnetic and electronic phases is well documented,\cite{park2021electronic,PhysRevB.87.115135,kiesel2013unconventional,jovanovic2022simple} though the intentional alignment of the Fermi level with the purported features in the band diagram has always been a challenging task. The synergistic doping of \kagSS~ is an excellent example of how control over electron filling can not only stabilize a crystal structure, but also provide a natural means to tune \ef. Further, given the strong magnetic character of the broader \textit{AM}$_3$\textit{X}$_4$ family, the (Sb,Sn) flexibility may provide a convenient means to investigate the impact of electron tuning on the magnetic and electronic properties. 

Figure \ref{fig:SmBasic} presents a summary of the magnetic, electronic, and thermodynamic properties for the Sn-rich and Sb-rich endpoints of the \SmSS~ series. We begin our discussion by focusing on the Sb-rich compound (green). Figure \ref{fig:SmBasic}(a) demonstrates the field-cooled (FC) and zero-field-cooled (ZFC) magnetization curves for a single crystal with the field oriented parallel to the three principle axes. \SmSS~ exhibits a highly anisotropic magnetic response with a clear easy \textit{c}-axis magnetization. Qualitatively the magnetization resembles a ferromagnetic (FM) system, replete with FC/ZFC splitting. However, a weak shoulder can be observed immediately before onset of the ferromagnetic response, the purported FM transition is broader than normally expected, and the plateau of approximately 1.5~emu/mol/Oe is fairly weak. 

Heat capacity measurements (Figure \ref{fig:SmBasic}(b)) indicate a somewhat broad second-order peak which coincides with the magnetization onset around 10~K. Closer analysis of the heat capacity peak clearly indicates two transitions separated by $<$1~K (see inset). The centroids of these peaks correspond to the weak shoulder and ZFC/FC splitting temperature in the magnetization results. Due to the lack of a non-magnetic standard in this system (we were unable to make either LaTi$_3$(Sb,Sn)$_4$ or CaTi$_3$(Sb,Sn)$_4$) we do not have a quantitative analysis of the magnetic entropy released, though a conservative estimate using a simple polynomial subtraction estimates 80\% of Rln(2), which is consistent with the prior results for magnetic Sm$^{3+}$ in the SmTi$_3$Bi$_4$ bismuthide.\cite{ortiz2023evolution}

The coincidence of the split heat capacity peak and the magnetization features suggests that there are potentially two closely-interacting magnetic transitions. To investigate, we performed field-dependent magnetization measurements at two temperatures: (1) the base temperature of 2~K, and (2) an intermediate temperature of 9.7~K (immediately before the onset of ZFC/FC splitting). At 9.7~K, Figure \ref{fig:SmBasic}(c) reveals clearly antiferromagnetic (AFM) behavior with multiple spin-flip transitions. Upon cooling to 2~K, however, the system develops substantial coercivity. Together these effects explain both the weak peak (AFM cusp) and plateau (coercivity) in the temperature-dependent magnetization Figure \ref{fig:SmBasic}(a). This effect is also observed in the symmetrized, transverse magnetoresistance (Figure \ref{fig:SmBasic}(d)), where we can observe the sharp onset of negative MR with a critical field that matches the second spin-flop transition around 0.9~T. As the system continues to cool we observe the evolution of the same hysteresis observed in the magnetization results.

We now turn to the Sn-rich endpoint. As established in the ARPES and DFT data, these compositions differ by approximately 0.5~Sn per formula unit (or approximately 0.5~holes per formula unit), with only a very weak (0.3\% change) response in the crystallographic parameters and minimal changes to the electronic structure. We anticipate that modifications to the magnetic ground state are primarily induced by changes in the electronic filling due to the synergistic doping effect. Figure \ref{fig:SmBasic}(e) presents the ZFC/FC magnetization results, again highlighting a highly anisotropic FM response with easy \textit{c}-axis magnetization. The Sn-rich compound exhibits a much sharper ferromagnetic onset and a plateau more in accordance with a prototypical FM. The broad peak from the Sb-rich compound reappears in the Sn-rich composition as well, though the separation is greater ($\approx$6~K above) and the cusp shape is much clearer. Naively, these results suggest an AFM ground state from approximately 15--20~K before the AFM state is overtaken by a stronger FM state. 

Heat capacity measurements (Figure \ref{fig:SmBasic}(f)) clearly show two distinct thermodynamic signatures corresponding to the AFM and FM transitions. The AFM transition (21~K) is second-order, but the AFM to FM peak (15~K) is clearly first-order in nature. The inset of Figure \ref{fig:SmBasic}(f) demonstrates the clear temperature splitting of the heating and cooling curves as measured by high-pulse single-slope heat capacity measurements. Isothermal magnetization measurements performed at temperatures between the two heat capacity peaks (Figure \ref{fig:SmBasic}(g)) confirms the onset of AFM order below 21~K, suppression of the AFM order by 18.5~K, and emergence of strong FM behavior by 2~K. Note that the saturation magnetization is not modified substantially between the Sb-rich and Sn-rich compositions, suggesting that the (Sb,Sn) is modifying the magnetic interactions, but not the effective moment or Crystalline Electric Fields (CEF) of the Sm ion. 

The electronic transport data for the Sn-rich composition (Figure \ref{fig:SmBasic}(h)) supports a similar interpretation of the data. As we cool from the paramagnetic state into the antiferromagnetic state, we see the Sn-rich sample begin to exhibit a similar negative magnetoresistance as in the Sb-rich sample. However, instead of the negative magnetoresistance developing the coercivity seen in the hybridized AFM/FM ground state of the Sb-rich (green) data, the Sn-rich sample (blue) shows the negative magnetoresistance vanish at the lower temperature phase transition. Furthermore, the critical field for the spin-flip transition visibly decreases as we continue cooling within the AFM state before fully disappearing in the FM states. This is strong evidence for the competition between AFM and FM states. Further, since the FM and AFM orders break different symmetries with different ordering vectors, the transition has to be first order (consistent with heat capacity results) since there is no symmetry allowed continuous route for this transition.

Let's broadly compare the Sb-rich and Sn-rich compositions. The Sn-rich (blue) composition exhibits a competition between AFM and FM order. The AFM order emerges first but is ultimately suppressed and replaced by the FM order. Contrast this with the Sb-rich (green) composition. After emergence of AFM order around 10-11~K, continued cooling sees substantial FM-like interactions emerge, indicated by both the irreversibility (coercivity) of the magnetization loops and the transport data. At this time, we suspect that the ground state of the system is likely either (1) AFM with substantial spin canting, (2) a ferrimagnet, or (3) AFM with evolution of a spin-density wave. At this time we have insufficient data to definitively characterize the magnetic nature of the Sb-rich \SmSS. For brevity we will refer to this state as an A(FM) to represent the interplay between the AFM character and the clear coercivity (FM interactions). Although both the Sb-rich and Sn-rich samples show both AFM and FM character at some point, we find it remarkable that (Sb,Sn) not only tunes the interaction strength, but also fundamentally changes the nature of the interaction. The Sb-rich compound shows a more seamless blending of the AFM and FM states into the A(FM) state, whereas the Sn-rich data shows a more stark competition (e.g. first order AFM to FM) between energy scales.

Thus far we have only shown the Sb-rich and Sn-rich termini of the \SmSS~ solid solution, however, seeing the evolution of the magnetic and thermodynamic properties as a function of composition further illustrates the interplay between the FM and AFM states. Figure \ref{fig:SmGrid} shows a series of \SmSS~ single crystals ranging between the Sb-rich and Sn-rich termini. From left to right, we display (1) the explicit (Sb,Sn) compositions as determined by energy dispersive spectroscopy (EDS), (2) temperature-dependent magnetization superimposed with zero-field heat capacity results, and (3) isothermal magnetization curves at base temperature (color) and an intermediate temperature (gray) that straddles the AFM and FM transitions. For graphical simplicity, all results are shown in arbitrarily scaled units, though a corresponding set of quantitative plots is provided in the supplemental information. 

Starting at the Sn-rich composition (Sample-I), we see the same highly coercive ferromagnetic loop and intermediate AFM-like state from Figure \ref{fig:SmBasic}(g). As additional Sb is incorporated into the structure, the heat capacity peaks begin to merge and the temperature-dependent magnetization loses the characteristic ``sharpness'' of the FM transition (e.g. Sample-III). Note that by Sample-III and Sample-IV, we can no longer identify a temperature where the intermediate temperature traces (gray) exhibit intermediate spin-flip plateaus. By Sample-V, the 2~K FM loop develops a characteristic rounding and the intermediate temperature begins to redevelop the spin-flip plateau. This is most evident as we transition from Sample-V to Sample-VI, where the 2~K loop fractures into the individual plateaus and shows clear blending of the AFM and FM states. By sample-VII the hybrid A(FM) state is clearly developed and the individual plateaus are sharp and well-defined. Together, these results clearly demonstrate that both temperature and composition (electron filling) tuning can be used to smoothly tune the system between the AFM, FM, and (A)FM states.

Figure \ref{fig:SmPhaseDia}(a) and (b) provide the temperature-composition phase diagram for \SmSS~ as derived from heat capacity and magnetization measurements. Heat capacity measurements (a) are likely the most detailed, as our high-pulse measurements can help identify both peak splittings and the difference between first- and second-order transitions. The inset helps indicate the color scheme for the samples, with the blue peaks indicating the highest (AFM) peak, and the green peaks indicating the lower-temperature first-order (FM) peaks. As demonstrated in Figure \ref{fig:SmBasic}(a, inset), the first-order peak always appears as a split peak with at least two distinct peaks. Both of the peak temperatures are highlighted on the plot. Similarly, after the initial peak merger shown in Figure \ref{fig:SmGrid}(Sample-III), the heat capacity peak broadens into a peak with a clear shoulder. While these peaks move together, they are distinct (as will be seen in the field-dependence shortly) and are marked as such here. 

The magnetization results can be slightly harder to interpret, largely because the strength of the FM transition often obscures the AFM cusp. Unfortunately, due to the myriad of other features that also emerge in the intermediate composition regime (recall Sample-V and Sample-VI from Figure \ref{fig:SmGrid}), using the temperature-derivative is also a challenging task. The inset of Figure \ref{fig:SmPhaseDia}(b) shows a small log-scale schematic of the two features which are present in almost all temperature-dependent magnetization measurements. There is commonly a small, weak cusp a higher temperatures (consistent with the AFM heat capacity peak) and then the much stronger FM transition. If we define T$_\text{AFM}$ as the cusp peak from the AFM transition and  T$_\text{FM}$ as the midpoint of the FM transition, then we can generate the magnetization-derived phase diagram shown in Figure \ref{fig:SmPhaseDia}(b). Both Figure \ref{fig:SmPhaseDia}(a) and Figure \ref{fig:SmPhaseDia}(b) agree qualitatively and quantitatively on the boundaries in the composition-temperature space. 

Figure \ref{fig:SmPhaseDia}(c-h) turn to examine the field-dependence of the Sb-rich and Sn-rich endpoints. This is particularly instructive when examining the heat capacity results, as we can clearly see the multiple peak structure of the results. As with the composition-temperature phase diagram, we will examine both the magnetization and heat capacity results to help provide a more exhaustive interpretation of the results. Figure \ref{fig:SmPhaseDia}(c,d) examine the field-dependence of the magnetization curves on a logarithmic scale. For the Sn-rich sample (c) The higher-temperature AFM feature is clearly visible for a relatively wide range of fields, though it is progressively suppressed with the application of fields. For the Sb-rich sample (d) the situation is a bit more complex. At low fields there is a clear shoulder at higher temperatures ($\approx$12~K) indicative of the onset of AFM-dominated interactions. However, unlike the Sn-rich sample, the application of a higher magnetic field doesn't progressively suppress the AFM interaction. Instead, there is an intermediate regime (note the black curve) where the AFM-like cusp is the dominant feature in the magnetization. This is consistent with the step-like features in the Sb-rich magnetization from Figure \ref{fig:SmBasic}(c), as we know that intermediate fields at approximately 10~K should be consistent with a magnetization response that resembles a prototypical spin-flop AFM. 

The heat-capacity curves are perhaps the most useful way to exhibit the interplay between the FM and AFM interactions in the system. Figure \ref{fig:SmPhaseDia}(e) shows the field-dependent (H$\parallel$c) heat capacity results for the Sn-rich terminus. The purple (zero-field) curve clearly shows the split first-order peak and the higher-temperature AFM-like second-order peak. We've shown the series of field-dependent measurements with every 5th curve highlighted for clarity. At small to moderate fields (0-1~T) the first-order FM peak is affected the most, shifting towards higher temperatures while the AFM peak is only slightly broadened/shifted to lower temperatures. Above 1~T the AFM peak is rapidly suppressed and shifted downwards, merging into the first-order peak, which has also broadened enough to lose the double peak structure. As suggested by all previous measurements, this indicates that the Sn-rich sample exhibits a higher-temperature AFM which competes (and loses) to a FM-dominated ground state. 

Now we turn to the more complex Sb-rich \SmSS~ field-dependent heat capacity in Figure \ref{fig:SmPhaseDia}(f). The zero-field (purple) curve is clearly broad, with a clear right and left shoulder. With small applied fields (up to about 0.2~T) the centroid of the peak shifts towards higher temperatures (FM-dominant), the right shoulder sharpens into a well-defined peak, and the left shoulder is suppressed/merged. Above 0.2~T the sharpened peak begins to shift towards lower temperatures (AFM-dominant). Close to $\approx$1~T, the peak quickly broadens into a double-peak feature right before the entire feature is suppressed in the field-polarized state. Again, we see the dance between AFM and FM interactions indicative of the \SmSS~ family, and the much more intertwined nature of the A(FM) ground state.

Figure \ref{fig:SmPhaseDia}(g,h) summarize the field-dependent magnetization and field-dependent heat capacity results as field-temperature phase diagrams. The background color contour is produced through the second-derivative of the magnetization curves, while the discrete points are derived from the field-dependent heat capacity results. Note that the heat capacity points in Figure \ref{fig:SmPhaseDia}(h) are determined as the centroid of the heat capacity peaks shown in Figure \ref{fig:SmPhaseDia}(f). 

Our results highlight how the intrinsic tunability inherited from the (Sb,Sn) solid-solution imparts a complex and nuanced degree of freedom to the \SmSS~ family of kagome metals. Not only does the (Sb,Sn) synergistic doping stabilize the crystal structure through controlling the electron filling and the relevance of bonding/antibonding states, but we also gain the ability to tune and control the interplay between the AFM and FM interactions intrinsic to the system. Particularly within the kagome metals, where the alignment of the Fermi level with critical features in the electronic band diagram is critical, the complex interplay between composition, temperature, and field in \SmSS~ offers a unique perspective on materials development. The strategy of utilizing synergistic dopants to develop more complex intermetallic materials is intriguing, as it is possible we are missing whole swaths intermetallic families without corresponding ``pure'' endpoints.

\subsection{Brief Commentary on \textit{Ln}:(Ce, Pr, Nd, Gd)}

For brevity and focus, this manuscript focused primarily on the properties of \SmSS~ series. For completeness we will briefly comment on the properties for the other rare-earth containing \kagSS, with analogous characterization and details shown in the supplementary information.

\textbf{\NdSS}: The Nd-containing series shows properties remarkably similar to the \SmSS~ series. The Sn-rich terminus resembles a FM ground state with onset of magnetic order around 10~K (see Supplemental Figure X) while Sb-rich \NdSS~ is very similar to the A(FM) state in Sb-rich \SmSS, demonstrating field-induced metamagnetic transitions with appreciable coercivity. Note that the magnetic anisotropy has changed, however, as \NdSS~ compounds exhibits easy \textit{b}-axis magnetization.

\textbf{\GdSS}: The Gd-containing series exhibits substantially less dependence on the (Sb,Sn) composition despite having a relatively wide solubility range. Both the Sb-rich and Sn-rich compositions show onset of antiferromagnetic order around 10-15~K and a clear metamagnetic transition when fields are directed along either \textit{a} or \textit{c}. The saturation moment is consistent with the full Gd$^{3+}$ moment. These results suggest that this compound may have a non-collinear AFM structure or other complex anisotropy. 

\textbf{\CeSS}: The Ce-containing compound exhibits an AFM cusp at 10~K followed by a FM-like (easy \textit{b}-axis) response that saturates upon cooling below 3~K. Properties are broadly consistent with the magnetization behavior seen in the Sn-rich \SmSS~ series. Field-dependent measurements at 8~K confirm AFM (metamagnetic) nature of the 10~K peak.

\textbf{\PrSS}: The Pr-containing compound exhibits a broad magnetization rise below approximately 5.5~K. Perhaps \PrSS~ exhibits features similar to the rest of the rare-earth compounds (e.g. a broad AFM shoulder followed by a FM rise with easy \textit{b}-axis magnetization), but the details are not clear. The case of Pr is further complicated by the known sensitivity of the Pr-ion to local disorder and CEF effects.

While brief, we believe that this summary (alongside the characterization data in the supplementary) will be sufficient for others to peruse the critical features of the other \kagSS~ compounds and identify potential avenues for future research in those compounds.

\section{Conclusions}

In this work we presented the single-crystal synthesis and characterization of the \kagSS~ series, a family which forms only as a solid-solution between the hypothetical \textit{Ln}Ti$_3$Sn$_4$ and \textit{Ln}Ti$_3$Sb$_4$ endpoints. We began our discussion by synthesizing the full range of \kagSS~ (\textit{Ln}: Ce--Gd) single crystals. We demonstrate an extended solid-solution in \NdSS~, \GdSS~, and \SmSS along with controllable incorporation of (Sb,Sn) by tuning the composition of the flux. We hypothesized that the \kagSS~ series is electronically stabilized by the (Sb,Sn) alloying. We subsequently employed a combination of ARPES, DFT, and COHP measurements to show that the (Sb,Sn) pair tunes \ef~ without appreciably changing the bulk crystal structure (0.3\% volumetric expansion) or the bulk electronic structure. Furthermore, (Sb,Sn) alloying appears to be strongly favored through (1) minimization of the density-of-states at \ef, (2) population of unfilled bonding states when alloying Sb into hypothetical \textit{Ln}Ti$_3$Sn$_4$, and (3) depopulation of filled antibonding states when alloying Sn into hypothetical \textit{Ln}Ti$_3$Sb$_4$. These results culminate in the (Sb,Sn) pair serving not only as an effective means to tune \ef, but also a stabilizing force that stabilizes the \kagSS~ crystal structure without deleterious effects on the electronic structure -- an effect we have dubbed ``synergistic doping.''

The (Sb,Sn) solid-solution also presents a unique opportunity to control the underlying electronic and magnetic properties of kagome metals, which is a persistent challenge in the development of new kagome metals. We chose to focus on the \SmSS~ series for a detailed discussion of the (Sb,Sn) doping on the physical properties, demonstrating a complex, tunable interaction between competing AFM and FM ground states. In the Sn-rich sample, the AFM and FM transitions are separated by $\approx$10~K and the FM state competes with and eventually suppresses the AFM state. In the Sb-rich sample, however, the AFM and FM order appear to emerge nearly simultaneously into a state we called A(FM). At this point we cannot definitively determine the magnetic ground state of the Sb-rich composition, but it is very likely either an AFM structure with substantial spin canting, a ferrimagnet, or a candidate for a spin-density wave or other spin texture. While our discussion focused on the \SmSS~ series, we briefly comment on the (Sb,Sn) alloying in the other rare-earth compounds (Ce, Pr, Nd, Gd) and provide additional details in supplementary information. 

Ultimately this work accomplishes two goals: (1) to present a new family of cleavable, air-stable kagome metals with complex magnetic ground states and high potential for correlated phenomena, and (2) use the (Sb,Sn) alloying in \kagSS~ to understand and propose a potentially new way of searching for complex intermetallics through the application of ``synergistic doping.'' 
\section{Acknowledgments}

\textbf{Notice: This manuscript has been authored by UT-Battelle, LLC, under contract DE-AC05-00OR22725 with the US Department of Energy (DOE). The US government retains and the publisher, by accepting the article for publication, acknowledges that the US government retains a nonexclusive, paid-up, irrevocable, worldwide license to publish or reproduce the published form of this manuscript, or allow others to do so, for US government purposes. DOE will provide public access to these results of federally sponsored research in accordance with the DOE Public Access Plan (https://www.energy.gov/doe-public-access-plan).}

This work (sample discovery, single-crystal synthesis, bulk properties measurements, ARPES, manuscript authoring), was primarily supported by the U.S. Department of Energy (DOE), Office of Science, Basic Energy Sciences (BES), Materials Sciences and Engineering Division. M.N., M.S., and A.K.K. (ARPES) acknowledge support from the National Science Foundation, Division of Materials Research, under Award No. 2518800. G.D.S. (DFT) acknowledges support from the Laboratory Directed Research and Development Program of Oak Ridge National Laboratory, managed by UTBattelle, LLC, for the US Department of Energy. This research used resources of the Compute and Data Environment for Science (CADES) at the Oak Ridge National Laboratory, which is supported by the Office of Science of the U.S. Department of Energy under Contract No. DE-AC05-00OR22725. This research used resources of the Advanced Light Source, a U.S. Department of Energy Office of Science User Facility, under Contract No. DE-AC02-05CH11231. This research used resources at 21-ID-1 beamline of the National Synchrotron Light Source II, a US Department of Energy Office of Science User Facility operated for the DOE Office of Science by Brookhaven National Laboratory under contract no. DE-SC0012704. Preliminary single-crystal neutron diffraction measurements on TOPAZ at the Spallation Neutron Source, a U.S. Department of Energy Office of Science User Facility operated by Oak Ridge National Laboratory, were carried out under IPTS-35625.1.

\section{Supplementary Info.}
Supplementary ARPES data on the Sb-rich and Sn-rich endpoints; Supplementary ARPES data showing comparison with SmTi$_3$Bi$_4$; Full set of COHP simulations; full set of MvH, MvT, CvT curves for alloy series with quantitative axes; Supplementary characterization data for \NdSS, \GdSS, \CeSS, and \PrSS; Corresponding CIF files and crystallographic table for Sb-rich endpoints;

\bibliography{MagneticTunable134}

@PREAMBLE{
 "\providecommand{\noopsort}[1]{}" 
 # "\providecommand{\singleletter}[1]{#1}%" 
}

@article{wang2024topological,
  title={Topological quantum materials with kagome lattice},
  author={Wang, Qi and Lei, Hechang and Qi, Yanpeng and Felser, Claudia},
  journal={Accounts of Materials Research},
  volume={5},
  number={7},
  pages={786--796},
  year={2024},
  publisher={ACS Publications}
}

@article{jiang2023kagome,
  title={Kagome superconductors av3sb5 (a= k, rb, cs)},
  author={Jiang, Kun and Wu, Tao and Yin, Jia-Xin and Wang, Zhenyu and Hasan, M Zahid and Wilson, Stephen D and Chen, Xianhui and Hu, Jiangping},
  journal={National Science Review},
  volume={10},
  number={2},
  pages={nwac199},
  year={2023},
  publisher={Oxford University Press}
}

@article{neupert2022charge,
  title={Charge order and superconductivity in kagome materials},
  author={Neupert, Titus and Denner, M Michael and Yin, Jia-Xin and Thomale, Ronny and Hasan, M Zahid},
  journal={Nature Physics},
  volume={18},
  number={2},
  pages={137--143},
  year={2022},
  publisher={Nature Publishing Group UK London}
}

@article{wang2023quantum,
  title={Quantum states and intertwining phases in kagome materials},
  author={Wang, Yaojia and Wu, Heng and McCandless, Gregory T and Chan, Julia Y and Ali, Mazhar N},
  journal={Nature Reviews Physics},
  volume={5},
  number={11},
  pages={635--658},
  year={2023},
  publisher={Nature Publishing Group UK London}
}

@article{yin2022topological,
  title={Topological kagome magnets and superconductors},
  author={Yin, Jia-Xin and Lian, Biao and Hasan, M Zahid},
  journal={Nature},
  volume={612},
  number={7941},
  pages={647--657},
  year={2022},
  publisher={Nature Publishing Group UK London}
}

@article{di2026kagome,
  title={Kagome metals},
  author={Di Sante, Domenico and Neupert, Titus and Sangiovanni, Giorgio and Thomale, Ronny and Comin, Riccardo and Checkelsky, Joseph G and Zeljkovic, Ilija and Wilson, Stephen D},
  journal={Reviews of Modern Physics},
  volume={98},
  number={1},
  pages={015002},
  year={2026},
  publisher={APS}
}

@article{wilson2024v3sb5,
  title={A V3Sb5 kagome superconductors},
  author={Wilson, Stephen D and Ortiz, Brenden R},
  journal={Nature Reviews Materials},
  volume={9},
  number={6},
  pages={420--432},
  year={2024},
  publisher={Nature Publishing Group UK London}
}

@Article{Wen2024_UnconventionalCDW-FeGe,
  author    = {Wen, Xikai and Zhang, Yuqing and Li, Chenglin and Gui, Zhigang and Li, Yikang and Li, Yanjun and Wu, Xueliang and Wang, Aifeng and Yang, Pengtao and Wang, Bosen and Cheng, Jinguang and Wang, Yilin and Ying, Jianjun and Chen, Xianhui},
  journal   = {Physical Review Research},
  title     = {Unconventional charge density wave in a kagome lattice antiferromagnet FeGe},
  year      = {2024},
  issn      = {2643-1564},
  month     = aug,
  number    = {3},
  pages     = {033222},
  volume    = {6},
  doi       = {10.1103/physrevresearch.6.033222},
  publisher = {American Physical Society (APS)},
}

@Article{Yu2024_MagAndCorrelationsScV6Sn6,
  author    = {Yu, Tianye and Lai, Junwen and Liu, Xiangyang and Liu, Peitao and Chen, Xing-Qiu and Sun, Yan},
  journal   = {Physical Review B},
  title     = {{Magnetism and weak electronic correlations in the kagome metal ScV$_6$Sn$_6$}},
  year      = {2024},
  issn      = {2469-9969},
  month     = may,
  number    = {19},
  pages     = {195145},
  volume    = {109},
  doi       = {10.1103/physrevb.109.195145},
  publisher = {American Physical Society (APS)},
}

@Article{Cao2023_CompetingChargeOrderScV6Sn6,
  author    = {Cao, Saizheng and Xu, Chenchao and Fukui, Hiroshi and Manjo, Taishun and Dong, Ying and Shi, Ming and Liu, Yang and Cao, Chao and Song, Yu},
  journal   = {Nature Communications},
  title     = {{Competing charge-density wave instabilities in the kagome metal ScV$_6$Sn$_6$}},
  year      = {2023},
  issn      = {2041-1723},
  month     = nov, 
  number = {1},
  pages    = {7671},
  volume    = {14},
  doi       = {10.1038/s41467-023-43454-1},
  publisher = {Springer Science and Business Media LLC},
}

@Article{Korshunov2023_SofteningPhononScV6Sn6,
  author    = {Korshunov, A. and Hu, H. and Subires, D. and Jiang, Y. and Călugăru, D. and Feng, X. and Rajapitamahuni, A. and Yi, C. and Roychowdhury, S. and Vergniory, M. G. and Strempfer, J. and Shekhar, C. and Vescovo, E. and Chernyshov, D. and Said, A. H. and Bosak, A. and Felser, C. and Bernevig, B. Andrei and Blanco-Canosa, S.},
  journal   = {Nature Communications},
  title     = {{Softening of a flat phonon mode in the kagome ScV$_6$Sn$_6$}},
  year      = {2023},
  issn      = {2041-1723},
  month     = oct,
  number    = {1},
  pages = {6646},
  volume    = {14},
  doi       = {10.1038/s41467-023-42186-6},
  publisher = {Springer Science and Business Media LLC},
}

@Article{Chen2024_LongRangeGeDimerizationFeGe,
  author    = {Chen, Ziyuan and Wu, Xueliang and Zhou, Shiming and Zhang, Jiakang and Yin, Ruotong and Li, Yuanji and Li, Mingzhe and Gong, Jiashuo and He, Mingquan and Chai, Yisheng and Zhou, Xiaoyuan and Wang, Yilin and Wang, Aifeng and Yan, Ya-Jun and Feng, Dong-Lai},
  journal   = {Nature Communications},
  title     = {Discovery of a long-ranged charge order with 1/4 Ge1-dimerization in an antiferromagnetic Kagome metal},
  year      = {2024},
  issn      = {2041-1723},
  month     = jul,
  number    = {1},
  volume    = {15},
  pages = {6262},
  doi       = {10.1038/s41467-024-50661-x},
  publisher = {Springer Science and Business Media LLC},
}

@Article{Wang2023_EnhancedSpinPolarizationViaDimerizationFeGe,
  author    = {Wang, Yilin},
  journal   = {Physical Review Materials},
  title     = {Enhanced spin-polarization via partial Ge-dimerization as the driving force of the charge density wave in FeGe},
  year      = {2023},
  issn      = {2475-9953},
  month     = oct,
  number    = {10},
  pages     = {104006},
  volume    = {7},
  doi       = {10.1103/physrevmaterials.7.104006},
  publisher = {American Physical Society (APS)},
}

@Article{Liu2024_DrivingMechanismScV6Sn6,
  author    = {Liu, Shuyuan and Wang, Chongze and Yao, Shichang and Jia, Yu and Zhang, Zhenyu and Cho, Jun-Hyung},
  journal   = {Physical Review B},
  title     = {Driving mechanism and dynamic fluctuations of charge density waves in the kagome metal ScV6Sn6},
  year      = {2024},
  issn      = {2469-9969},
  month     = mar,
  number    = {12},
  pages     = {l121103},
  volume    = {109},
  doi       = {10.1103/physrevb.109.l121103},
  publisher = {American Physical Society (APS)},
}

@Article{Lee2024_NatureCDWScV6Sn6,
  author    = {Lee, Seongyong and Won, Choongjae and Kim, Jimin and Yoo, Jonggyu and Park, Sudong and Denlinger, Jonathan and Jozwiak, Chris and Bostwick, Aaron and Rotenberg, Eli and Comin, Riccardo and Kang, Mingu and Park, Jae-Hoon},
  journal   = {npj Quantum Materials},
  title     = {Nature of charge density wave in kagome metal ScV6Sn6},
  year      = {2024},
  issn      = {2397-4648},
  month     = jan,
  number    = {1},
  pages = {15},
  volume    = {9},
  doi       = {10.1038/s41535-024-00620-y},
  publisher = {Springer Science and Business Media LLC},
}

@Article{Hu2024_PhononPromotedCDW-ScV6Sn6,
  author    = {Hu, Yong and Ma, Junzhang and Li, Yinxiang and Jiang, Yuxiao and Gawryluk, Dariusz Jakub and Hu, Tianchen and Teyssier, Jérémie and Multian, Volodymyr and Yin, Zhouyi and Xu, Shuxiang and Shin, Soohyeon and Plokhikh, Igor and Han, Xinloong and Plumb, Nicholas C. and Liu, Yang and Yin, Jia-Xin and Guguchia, Zurab and Zhao, Yue and Schnyder, Andreas P. and Wu, Xianxin and Pomjakushina, Ekaterina and Hasan, M. Zahid and Wang, Nanlin and Shi, Ming},
  journal   = {Nature Communications},
  title     = {{Phonon promoted charge density wave in topological kagome metal ScV$_6$Sn$_6$}},
  year      = {2024},
  issn      = {2041-1723},
  month     = feb,
  pages    = {1658},
  volume    = {15},
  doi       = {10.1038/s41467-024-45859-y},
  publisher = {Springer Science and Business Media LLC},
}

@Misc{Hu2023_ScV6Sn6-Theory-FlatPhonons+UnconventionalCDW,
  author    = {Hu, Haoyu and Jiang, Yi and Călugăru, Dumitru and Feng, Xiaolong and Subires, David and Vergniory, Maia G. and Felser, Claudia and Blanco-Canosa, Santiago and Bernevig, B. Andrei},
  title     = {Kagome Materials I: SG 191, ScV$_6$Sn$_6$. Flat Phonon Soft Modes and Unconventional CDW Formation: Microscopic and Effective Theory},
  year      = {2023},
  copyright = {arXiv.org perpetual, non-exclusive license},
  doi       = {10.48550/ARXIV.2305.15469},
  publisher = {arXiv},
}

@Article{Pokharel2023_FrustratedCO+CooperativeDistortionsScV6Sn6,
  author    = {Pokharel, Ganesh and Ortiz, Brenden R. and Kautzsch, Linus and Gomez Alvarado, S. J. and Mallayya, Krishnanand and Wu, Guang and Kim, Eun-Ah and Ruff, Jacob P. C. and Sarker, Suchismita and Wilson, Stephen D.},
  journal   = {Physical Review Materials},
  title     = {{Frustrated charge order and cooperative distortions in ScV$_6$Sn$_6$}},
  year      = {2023},
  issn      = {2475-9953},
  month     = oct,
  number    = {10},
  pages     = {104201},
  volume    = {7},
  doi       = {10.1103/physrevmaterials.7.104201},
  publisher = {American Physical Society (APS)},
}

@misc{maintz2016lobster,
  title={LOBSTER: a tool to extract chemical bonding from plane-wave based DFT},
  author={Maintz, Stefan and Deringer, Volker L and Tchougr{\'e}eff, Andrei L and Dronskowski, Richard},
  year={2016},
  publisher={Wiley Online Library}
}

@article{wang2025loposter,
  title={LOPOSTER: a cascading postprocessor for LOBSTER},
  author={Wang, YiXu and M{\"u}ller, Peter C and Hemker, David and Dronskowski, Richard},
  journal={Journal of Computational Chemistry},
  volume={46},
  number={17},
  pages={e70167},
  year={2025},
  publisher={Wiley Online Library}
}

@article{maintz2013analytic,
  title={Analytic projection from plane-wave and PAW wavefunctions and application to chemical-bonding analysis in solids},
  author={Maintz, Stefan and Deringer, Volker L and Tchougr{\'e}eff, Andrei L and Dronskowski, Richard},
  journal={Journal of computational chemistry},
  volume={34},
  number={29},
  pages={2557--2567},
  year={2013},
  publisher={Wiley Online Library}
}

@article{COHP_deringer2011crystal,
  title={Crystal orbital Hamilton population (COHP) analysis as projected from plane-wave basis sets},
  author={Deringer, Volker L and Tchougr{\'e}eff, Andrei L and Dronskowski, Richard},
  journal={The journal of physical chemistry A},
  volume={115},
  number={21},
  pages={5461--5466},
  year={2011},
  publisher={ACS Publications}
}

@article{COHP_dronskowski1993crystal,
  title={Crystal orbital Hamilton populations (COHP): energy-resolved visualization of chemical bonding in solids based on density-functional calculations},
  author={Dronskowski, Richard and Bloechl, Peter E},
  journal={The Journal of Physical Chemistry},
  volume={97},
  number={33},
  pages={8617--8624},
  year={1993},
  publisher={ACS Publications}
}

@article{kohn1965self,
  title={Self-consistent equations including exchange and correlation effects},
  author={Kohn, Walter and Sham, Lu Jeu},
  journal={Physical review},
  volume={140},
  number={4A},
  pages={A1133},
  year={1965},
  publisher={APS}
}

@article{ortiz2025isolated,
  title={Isolated spin ladders in L n 2 Ti 9 Sb 11 (L n: La--Nd) metals},
  author={Ortiz, Brenden R and Zhang, Heda and G{\'o}rnicka, Karolina and Cook, Matthew S and Sarker, Suchismita and Okamoto, Satoshi and Yan, Jiaqiang},
  journal={Physical Review Materials},
  volume={9},
  number={8},
  pages={086203},
  year={2025},
  publisher={APS}
}

@article{coates2018suite,
  title={A suite-level review of the neutron single-crystal diffraction instruments at Oak Ridge National Laboratory},
  author={Coates, Leighton and Cao, HB and Chakoumakos, Bryan C and Frontzek, Matthias D and Hoffmann, Christina and Kovalevsky, Andrii Y and Liu, Yaohua and Meilleur, Flora and dos Santos, Antonio M and Myles, Dean AA and others},
  journal={Review of Scientific Instruments},
  volume={89},
  number={9},
  year={2018},
  publisher={AIP Publishing}
}

@article{EuXport_shu2025complex,
  title={Complex magnetotransport in the paramagnetic state of the magnetic kagome metal EuTi 3 Bi 4},
  author={Shu, Yun and Mi, Xinrun and Wei, Yuhao and Tao, Sixue and Wang, Aifeng and Chai, Yisheng and Ma, Dashuai and Yang, Xiaolong and He, Mingquan},
  journal={Physical Review B},
  volume={111},
  number={15},
  pages={155103},
  year={2025},
  publisher={APS}
}

@article{GdXport_li2025anisotropic,
  title={Anisotropic magnetoresistance in antiferromagnetic kagome metal GdTi3Bi4},
  author={Li, Xinyao and Yang, Yang and Guan, Feihong and Zhu, Xiangde and Ning, Wei and Tian, Mingliang},
  journal={Applied Physics Letters},
  volume={126},
  number={9},
  year={2025},
  publisher={AIP Publishing}
}

@article{YbOsc_shtefiienko2025electronic,
  title={Electronic structure of the kagome metal YbTi 3 Bi 4 studied using torque magnetometry},
  author={Shtefiienko, Kyryl and Phillips, Cole and Ortiz, Brenden R and Graf, David E and Shrestha, Keshav},
  journal={Physical Review B},
  volume={111},
  number={3},
  pages={035145},
  year={2025},
  publisher={APS}
}

@article{GdMFM_guo2025tunable,
  title={Tunable Bifurcation of Magnetic Anisotropy and Bi-Oriented Antiferromagnetic Order in Kagome Metal GdTi 3 Bi 4},
  author={Guo, Jianfeng and Zhu, Shiyu and Zhou, Runnong and Wang, Ruwen and Wang, Yunhao and Sun, Jianping and Zhao, Zhen and Dong, Xiaoli and Cheng, Jinguang and Yang, Haitao and others},
  journal={Physical Review Letters},
  volume={134},
  number={22},
  pages={226704},
  year={2025},
  publisher={APS}
}

@article{TbSTM_zhang2025observation,
  title={Observation of orbital-selective dual modulations in an anisotropic antiferromagnetic kagome metal TbTi 3 Bi 4},
  author={Zhang, Renjie and Yu, Bocheng and Tan, Hengxin and Cheng, Yiwei and Shen, Feiran and Yang, Junye and Mu, Dan and Han, Xinru and Zong, Alfred and Hu, Quanxin and others},
  journal={Physical Review X},
  volume={15},
  number={3},
  pages={031012},
  year={2025},
  publisher={APS}
}

@article{LaARP_sakhya2025diverse,
  title={Diverse electronic topography in a distorted kagome metal LaTi3Bi4},
  author={Sakhya, Anup Pradhan and Ortiz, Brenden R and Ghosh, Barun and Sprague, Milo and Mondal, Mazharul Islam and Matzelle, Matthew and Atlam, Nabil and Kumay, Arun K and Mandrus, David G and Denlinger, Jonathan D and others},
  journal={arXiv preprint arXiv:2503.15759},
  year={2025}
}

@article{YbARP_sakhya2024diverse,
  title={Diverse electronic landscape of the kagome metal YbTi3Bi4},
  author={Sakhya, Anup Pradhan and Ortiz, Brenden R and Ghosh, Barun and Sprague, Milo and Mondal, Mazharul Islam and Matzelle, Matthew and Bin Elius, Iftakhar and Valadez, Nathan and Mandrus, David G and Bansil, Arun and others},
  journal={Communications Materials},
  volume={5},
  number={1},
  pages={241},
  year={2024},
  publisher={Nature Publishing Group UK London}
}

@article{TbARP_kushnirenko2025observation,
  title={Observation of band splitting and magnetically induced band structure reconstruction in TbTi 3 Bi 4},
  author={Kushnirenko, Yevhen and Wang, Lin-Lin and Su, Xiaoyi and Eaton, Andrew and Canfield, PC and Kaminski, Adam and Schrunk, Benjamin and O'Leary, Evan},
  journal={Physical Review B},
  volume={112},
  number={15},
  pages={155119},
  year={2025},
  publisher={APS}
}

@article{TbARP_zhang2024observation,
  title={Observation of Orbital-Selective Band Reconstruction in an Anisotropic Antiferromagnetic Kagome Metal TbTi3Bi4},
  author={Zhang, Renjie and Yu, Bocheng and Tan, Hengxin and Cheng, Yiwei and Zong, Alfred and Hu, Quanxin and Chen, Xuezhi and Hu, Yudong and Meng, Chengnuo and Ren, Junchao and others},
  journal={arXiv preprint arXiv:2412.16815},
  year={2024}
}

@article{GdARP_cheng2024striped,
  title={Striped magnetization plateau and chirality-reversible anomalous Hall effect in a magnetic kagome metal},
  author={Cheng, Erjian and Mao, Ning and Yang, Xiaotian and Song, Boqing and Lou, Rui and Ying, Tianping and Nie, Simin and Fedorov, Alexander and Bertran, Fran{\c{c}}ois and Ding, Pengfei and others},
  journal={arXiv preprint arXiv:2409.01365},
  year={2024}
}

@article{EuARP_jiang2024topological,
  title={Topological surface states in quasi-two-dimensional magnetic kagome metal EuTi3Bi4},
  author={Jiang, Zhicheng and Li, Tongrui and Yuan, Jian and Liu, Zhengtai and Cao, Zhipeng and Cho, Soohyun and Shu, Mingfang and Yang, Yichen and Li, Zhikai and Liu, Jiayu and others},
  journal={Science bulletin},
  volume={69},
  number={20},
  pages={3192--3196},
  year={2024}
}

@article{SmARP_zheng2024anisotropic,
  title={Anisotropic magnetism and band evolution induced by ferromagnetic phase transition in titanium-based kagome ferromagnet SmTi3Bi4},
  author={Zheng, Zhe and Chen, Long and Ji, Xuecong and Zhou, Ying and Qu, Gexing and Hu, Mingzhe and Huang, Yaobo and Weng, Hongming and Qian, Tian and Wang, Gang},
  journal={Science China Physics, Mechanics \& Astronomy},
  volume={67},
  number={6},
  pages={267411},
  year={2024},
  publisher={Springer}
}

@article{NdARP_hu2024magnetic,
  title={Magnetic coupled electronic landscape in bilayer-distorted titanium-based kagome metals},
  author={Hu, Yong and Le, Congcong and Chen, Long and Deng, Hanbin and Zhou, Ying and Plumb, Nicholas C and Radovic, Milan and Thomale, Ronny and Schnyder, Andreas P and Yin, Jia-Xin and others},
  journal={Physical Review B},
  volume={110},
  number={12},
  pages={L121114},
  year={2024},
  publisher={APS}
}

@article{NdARP_mondal2025observation,
  title={Observation of multiple flat bands and Van Hove singularities in the distorted kagome metal NdTi 3 Bi 4},
  author={Mondal, Mazharul Islam and Sakhya, Anup Pradhan and Sprague, Milo and Ortiz, Brenden R and Matzelle, Matthew and Kumay, Arun K and Seal, Avike and Ghosh, Barun and Bansil, Arun and Neupane, Madhab},
  journal={Physical Review B},
  volume={112},
  number={12},
  pages={L121104},
  year={2025},
  publisher={APS}
}

@article{CeCDW_park2025spin,
  title={Spin density wave and van Hove singularity in the kagome metal CeTi3Bi4},
  author={Park, Pyeongjae and Ortiz, Brenden R and Sprague, Milo and Sakhya, Anup Pradhan and Chen, Si Athena and Frontzek, Matthias D and Tian, Wei and Sibille, Romain and Mazzone, Daniel G and Tabata, Chihiro and others},
  journal={Nature Communications},
  volume={16},
  number={1},
  pages={4384},
  year={2025},
  publisher={Nature Publishing Group UK London}
}

@article{GdCDW_han2025discovery,
  title={Discovery of unconventional charge-spin-intertwined density wave in magnetic kagome metal GdTi3Bi4},
  author={Han, Xianghe and Chen, Hui and Cao, Zhongyi and Guo, Jingwen and Fei, Fucong and Tan, Hengxin and Guo, Jianfeng and Shi, Yanhao and Zhou, Runnong and Wang, Ruwen and others},
  journal={arXiv preprint arXiv:2503.05545},
  year={2025}
}

@article{TbCDW_cheng2024spectroscopic,
  title={Spectroscopic origin of giant anomalous Hall effect in an interwoven magnetic kagome metal},
  author={Cheng, Erjian and Wang, Kaipu and Hao, Yiqing and Chen, Wenqing and Tan, Hengxin and Li, Zongkai and Wang, Meixiao and Gao, Wenli and Wu, Di and Sun, Shuaishuai and others},
  journal={arXiv preprint arXiv:2405.16831},
  year={2024}
}

@article{arachchige2022charge,
  title={Charge density wave in kagome lattice intermetallic ScV 6 Sn 6},
  author={Arachchige, Hasitha W Suriya and Meier, William R and Marshall, Madalynn and Matsuoka, Takahiro and Xue, Rui and McGuire, Michael A and Hermann, Raphael P and Cao, Huibo and Mandrus, David},
  journal={Physical Review Letters},
  volume={129},
  number={21},
  pages={216402},
  year={2022},
  publisher={APS}
}

@article{ortiz2025stability,
  title={Stability Frontiers in the AM 6 X 6 Kagome Metals: The Ln Nb6Sn6 (Ln: Ce--Lu, Y) Family and Density-Wave Transition in LuNb6Sn6},
  author={Ortiz, Brenden R and Meier, William R and Pokharel, Ganesh and Chamorro, Juan and Yang, Fazhi and Mozaffari, Shirin and Thaler, Alex and Gomez Alvarado, Steven J and Zhang, Heda and Parker, David S and others},
  journal={Journal of the American Chemical Society},
  volume={147},
  number={6},
  pages={5279--5292},
  year={2025},
  publisher={ACS Publications}
}

@article{meier2023tiny,
  title={Tiny Sc allows the chains to rattle: impact of Lu and Y Doping on the charge-density wave in ScV6Sn6},
  author={Meier, William R and Madhogaria, Richa Pokharel and Mozaffari, Shirin and Marshall, Madalynn and Graf, David E and McGuire, Michael A and Arachchige, Hasitha W Suriya and Allen, Caleb L and Driver, Jeremy and Cao, Huibo and others},
  journal={Journal of the American Chemical Society},
  volume={145},
  number={38},
  pages={20943--20950},
  year={2023},
  publisher={ACS Publications}
}

@article{ortiz2024intricate,
  title={Intricate magnetic landscape in antiferromagnetic kagome metal TbTi3Bi4 and interplay with Ln2--x Ti6+ x Bi9 (Ln: Tb{\textperiodcentered}{\textperiodcentered}{\textperiodcentered} Lu) shurikagome metals},
  author={Ortiz, Brenden R and Zhang, Heda and G{\'o}rnicka, Karolina and Parker, David S and Samolyuk, German D and Yang, Fazhi and Miao, Hu and Lu, Qiangsheng and Moore, Robert G and May, Andrew F and others},
  journal={Chemistry of Materials},
  volume={36},
  number={16},
  pages={8002--8014},
  year={2024},
  publisher={ACS Publications}
}

@article{jovanovic2022simple,
  title={{Simple chemical rules for predicting band structures of kagome materials}},
  author={Jovanovic, Milena and Schoop, Leslie M},
  journal={Journal of the American Chemical Society},
  volume={144},
  number={24},
  pages={10978--10991},
  year={2022},
  publisher={ACS Publications}
}

@article{ortiz2023evolution,
  title={{Evolution of Highly Anisotropic Magnetism in the Titanium-Based Kagome Metals LnTi$_3$Bi$_4$ (Ln: La{\textperiodcentered}{\textperiodcentered}{\textperiodcentered} Gd$^{3+}$, Eu$^{2+}$, Yb$^{2+}$)}},
  author={Ortiz, Brenden R and Miao, Hu and Parker, David S and Yang, Fazhi and Samolyuk, German D and Clements, Eleanor M and Rajapitamahuni, Anil and Yilmaz, Turgut and Vescovo, Elio and Yan, Jiaqiang and others},
  journal={Chemistry of Materials},
  volume={35},
  number={22},
  pages={9756--9773},
  year={2023},
  publisher={ACS Publications}
}

@article{liu2023superconductivity,
  title={{Superconductivity emerged from density-wave order in a kagome bad metal}},
  author={Liu, Yi and Liu, Zi-Yi and Bao, Jin-Ke and Yang, Peng-Tao and Ji, Liang-Wen and Liu, Ji-Yong and Xu, Chen-Chao and Yang, Wu-Zhang and Chai, Wan-Li and Lu, Jia-Yi and others},
  journal={\textbf{2023}, arXiv:2309.13514 [cond-mat.supr-con]. arXiv.org e-Print archive. //https://arxiv.org/pdf/2309.13514 (Accessed 6-25-2024)}
}

@article{perdew1996,
  title={Generalized gradient approximation made simple},
  author={Perdew, John P and Burke, Kieron and Ernzerhof, Matthias},
  journal={Phys. Rev. Lett.},
  volume={77},
  number={18},
  pages={3865},
  year={1996},
  publisher={APS}

}

@article{canfield2016use,
  title={Use of frit-disc crucibles for routine and exploratory solution growth of single crystalline samples},
  author={Canfield, Paul C and Kong, Tai and Kaluarachchi, Udhara S and Jo, Na Hyun},
  journal={Philosophical magazine},
  volume={96},
  number={1},
  pages={84--92},
  year={2016},
  publisher={Taylor \& Francis}
}

@article{guo2023134,
  title={{Magnetic kagome materials RETi$_3$Bi$_4$ family with weak interlayer interactions}},
  author={Guo, Jingwen and Zhou, Liqin and Ding, Jianyang and Qu, Gexing and Liu, Zhengtai and Du, Yu and Zhang, Heng and Li, Jiajun and Zhang, Yiying and Zhou, Fuwei and Qi, Wuyi and Guo, Fengyi and Wang, Tianqi and Fei, Fucong and Huang, Yaobo and Qian, Tian and Shen, Dawei and Weng, Hongming and Song, Fengqi},
  journal={\textbf{2023}, arXiv:2308.14509v1 [cond-mat.mtrl-sci]. arXiv.org e-Print archive. https://arxiv.org/pdf/2308.14509.pdf (Accessed 10-24-2023)}
}

@article{chen2023134,
  title={Tunable magnetism in titanium-based kagome metals by rare-earth engineering and high pressure},
  author={Chen, Long and Zhou, Ying and Zhang, He and Ji, Xuecong and Liao, Ke and Ji, Yu and Li, Ying and Guo, Zhongnan and Shen, Xi and Yu, Richeng and others},
  journal={Communications Materials},
  volume={5},
  number={1},
  pages={73},
  year={2024},
  publisher={Nature Publishing Group UK London}
}

@article{ortiz2023ybv,
  title={{YbV$_3$Sb$_4$ and EuV$_3$Sb$_4$ vanadium-based kagome metals with Yb$^{2+}$ and Eu$^{2+}$ zigzag chains}},
  author={Ortiz, Brenden R and Pokharel, Ganesh and Gundayao, Malia and Li, Hong and Kaboudvand, Farnaz and Kautzsch, Linus and Sarker, Suchismita and Ruff, Jacob PC and Hogan, Tom and Alvarado, Steven J Gomez and others},
  journal={Phys. Rev. Mater.},
  volume={7},
  number={6},
  pages={064201},
  year={2023},
  publisher={APS}
}

@article{teng2022discovery,
  title={Discovery of charge density wave in a kagome lattice antiferromagnet},
  author={Teng, Xiaokun and Chen, Lebing and Ye, Feng and Rosenberg, Elliott and Liu, Zhaoyu and Yin, Jia-Xin and Jiang, Yu-Xiao and Oh, Ji Seop and Hasan, M Zahid and Neubauer, Kelly J and others},
  journal={Nature},
  volume={609},
  number={7927},
  pages={490--495},
  year={2022},
  publisher={Nature Publishing Group UK London}
}

@article{kresse1996efficiency,
  title={Efficiency of ab-initio total energy calculations for metals and semiconductors using a plane-wave basis set},
  author={Kresse, Georg and Furthm{\"u}ller, J{\"u}rgen},
  journal={Comput. Mater. Sci.},
  volume={6},
  number={1},
  pages={15--50},
  year={1996},
  publisher={Elsevier}
}

@article{blochl1994projector,
  title={Projector augmented-wave method},
  author={Bl{\"o}chl, Peter E},
  journal={Phys. Rev. B},
  volume={50},
  number={24},
  pages={17953},
  year={1994},
  publisher={APS}
}

@article{kresse1999ultrasoft,
  title={From ultrasoft pseudopotentials to the projector augmented-wave method},
  author={Kresse, Georg and Joubert, Daniel},
  journal={Phys. Rev. B},
  volume={59},
  number={3},
  pages={1758},
  year={1999},
  publisher={APS}
}

@article{ovchinnikov2018synthesis,
  title={{Synthesis, Crystal and Electronic Structure of the Titanium Bismuthides Sr$_5$Ti$_{12}$Bi$_{19+x}$, Ba$_5$Ti$_{12}$Bi$_{19+x}$, and Sr$_{5-\delta}$Eu$_\delta$Ti$_{12}$Bi$_{19+x}$ (x=0.5--1.0; $\delta$=2.4, 4.0)}},
  author={Ovchinnikov, Alexander and Bobev, Svilen},
  journal={Eur. J. Inorg. Chem.},
  volume={2018},
  number={11},
  pages={1266--1274},
  year={2018},
  publisher={Wiley Online Library}
}

@article{werhahn2022kagome,
  title={{The kagom{\'e} metals RbTi$_3$Bi$_5$ and CsTi$_3$Bi$_5$}},
  author={Werhahn, Dominik and Ortiz, Brenden R and Hay, Aurland K and Wilson, Stephen D and Seshadri, Ram and Johrendt, Dirk},
  journal={Z. Naturforsch. B},
  volume={77},
  number={11-12},
  pages={757--764},
  year={2022},
  publisher={De Gruyter}
}

@article{motoyama2018magnetic,
  title={{Magnetic properties of new antiferromagnetic heavy-fermion compounds, Ce$_3$TiBi$_5$ and CeTi$_3$Bi$_4$}},
  author={Motoyama, Gaku and Sezaki, Masumi and Gouchi, Jun and Miyoshi, Kiyotaka and Nishigori, Shijo and Mutou, Tetsuya and Fujiwara, Kenji and Uwatoko, Yoshiya},
  journal={Physica B Condens.},
  volume={536},
  pages={142--144},
  year={2018},
  publisher={Elsevier}
}

@article{bie2007ternary,
  title={{Ternary rare-earth titanium antimonides: phase equilibria in the RE--Ti--Sb (RE= La, Er) systems and crystal structures of RE$_2$Ti$_7$Sb$_{12}$ (RE= La, Ce, Pr, Nd) and RETi$_3$(Sn$_x$Sb$_{1-x}$)$_4$ (RE= Nd, Sm)}},
  author={Bie, Haiying and Moore, SH Devon and Piercey, Davin G and Tkachuk, Andriy V and Zelinska, Oksana Ya and Mar, Arthur},
  journal={J. Solid State Chem."},
  volume={180},
  number={8},
  pages={2216--2224},
  year={2007},
  publisher={Elsevier}
}

@article{ovchinnikov2019bismuth,
  title={Bismuth as a reactive solvent in the synthesis of multicomponent transition-metal-bearing bismuthides},
  author={Ovchinnikov, Alexander and Bobev, Svilen},
  journal={Inorg. Chem.},
  volume={59},
  number={6},
  pages={3459--3470},
  year={2019},
  publisher={ACS Publications}
}

@article{PhysRevB.87.115135,
	title = {Competing electronic orders on kagome lattices at van Hove filling},
	author = {Wang, Wan-Sheng and Li, Zheng-Zhao and Xiang, Yuan-Yuan and Wang, Qiang-Hua},
	journal = {Phys. Rev. B},
	volume = {87},
	issue = {11},
	pages = {115135},
	numpages = {8},
	year = {2013},
	month = {Mar},
	publisher = {American Physical Society},
	doi = {10.1103/PhysRevB.87.115135},
	url = {https://link.aps.org/doi/10.1103/PhysRevB.87.115135}
}

@article{kiesel2013unconventional,
  title={{Unconventional Fermi surface instabilities in the kagome Hubbard model}},
  author={Kiesel, Maximilian L and Platt, Christian and Thomale, Ronny},
  journal={Phys. Rev. Lett.},
  volume={110},
  number={12},
  pages={126405},
  year={2013},
  publisher={APS}
}

@article{ortiz2019new,
  title={{New kagome prototype materials: discovery of KV$_3$Sb$_5$, RbV$_3$Sb$_5$, and CsV$_3$Sb$_5$}},
  author={Ortiz, Brenden R and Gomes, L{\'\i}dia C and Morey, Jennifer R and Winiarski, Michal and Bordelon, Mitchell and Mangum, John S and Oswald, Iain WH and Rodriguez-Rivera, Jose A and Neilson, James R and Wilson, Stephen D and others},
  journal={Phys. Rev. Materials},
  volume={3},
  number={9},
  pages={094407},
  year={2019},
  publisher={APS}
}

@article{ortizCsV3Sb5,
  title={{CsV$_3$Sb$_5$: a $Z_2$ topological kagome metal with a superconducting ground state}},
  author={Ortiz, Brenden R and Teicher, Samuel M.L. and Hu, Yong and Zuo, Julia L and Sarte, Paul M. and Schueller, Emily C and Abeykoon, A.M. Milinda and Krogstad, Michael J and Rosenkranz, Stefan and Osborn, Raymond and Seshadri, Ram and Balents, Leon and He, Jungfeng and Wilson, Stephen D},
  journal={Phys. Rev. Lett.},
  volume={125},
  number={24},
  pages={247002},
  year={2020}
}

@article{park2021electronic,
  title={{Electronic instabilities of kagome metals: saddle points and Landau theory}},
  author={Park, Takamori and Ye, Mengxing and Balents, Leon},
  journal={Phys. Rev. B},
  volume={104},
  number={3},
  pages={035142},
  year={2021},
  publisher={APS}
}

\clearpage
\onecolumngrid

\pagestyle{empty}

\setcounter{figure}{0}
\renewcommand{\thefigure}{S\arabic{figure}}

\clearpage
\thispagestyle{empty}
\vspace*{\fill}
\begin{figure}[H]
\centering
\includegraphics[width=1\textwidth]{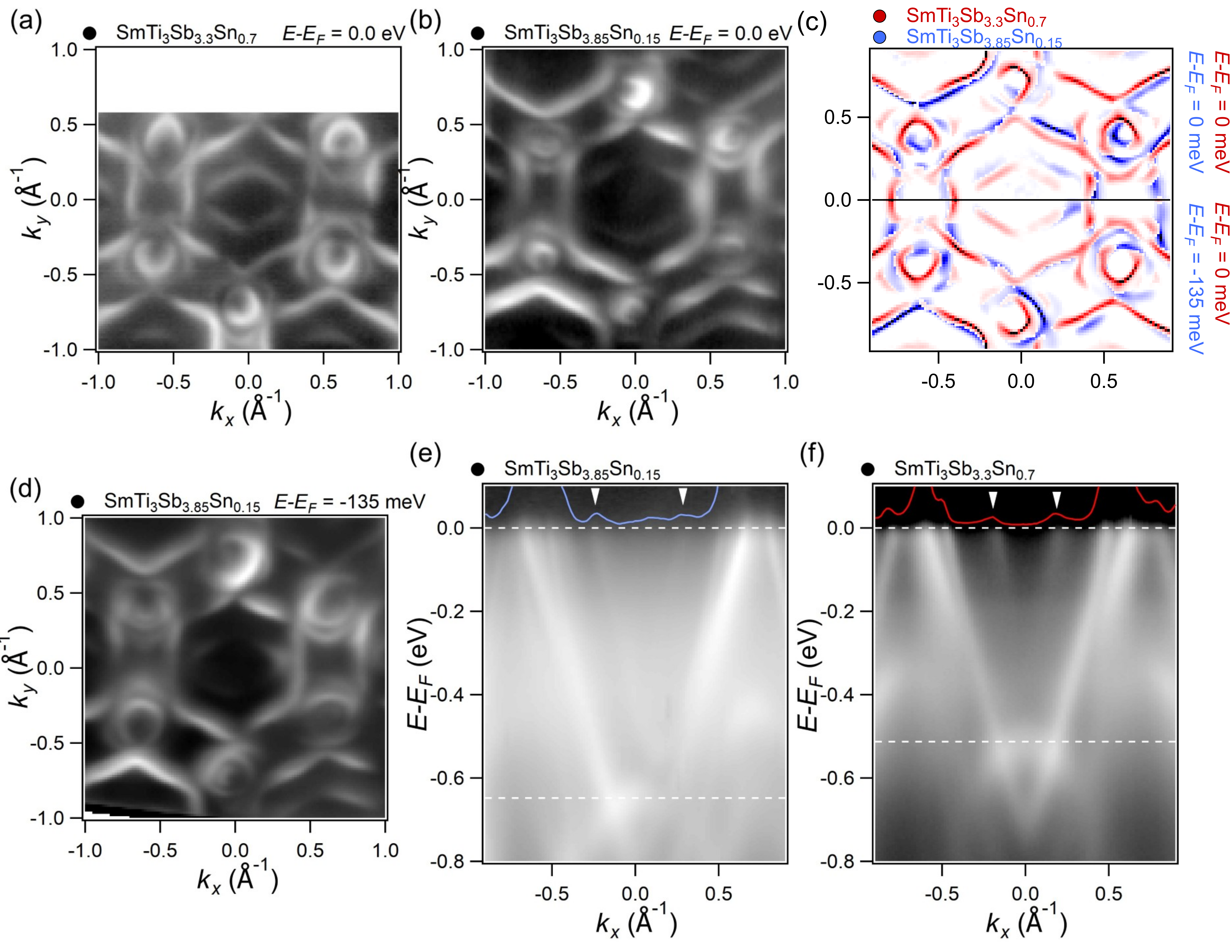}
\caption{Supplementary ARPES data collected on the Sn-Rich and Sb-Rich \SmSS~ series. (a) Constant energy contour (CEC) for Sb-rich \SmSS~ at the natural Fermi level. (b) CEC for Sn-rich \SmSS~ at the natural Fermi level. While the Sb-rich and Sn-rich compounds look broadly similar, there is a clear shift in the bands near $\Gamma$. (c) Comparison between the Sn-rich and Sb-rich ARPES scans at the natural Fermi level (top) and the same materials but with the Sb-rich CEC shifted to the -135~meV level. We can see that the bands near $\Gamma$ are much more aligned with the 135~meV shift. This shift corresponds to the doping difference between the Sb-rich and Sn-rich \SmSS. (d) CEC of the Sb-rich compound at the -135~meV Fermi level. (e,f) Same ARPES data shown in the main body, highlighting the bands near $\Gamma$ and the Fermi level shift induced by the Sb-Sn doping.}
\label{fig:FLarpes}
\end{figure}
\vspace*{\fill}
\clearpage

\clearpage
\thispagestyle{empty}
\vspace*{\fill}
\begin{figure}[H]
\centering
\includegraphics[width=1\textwidth]{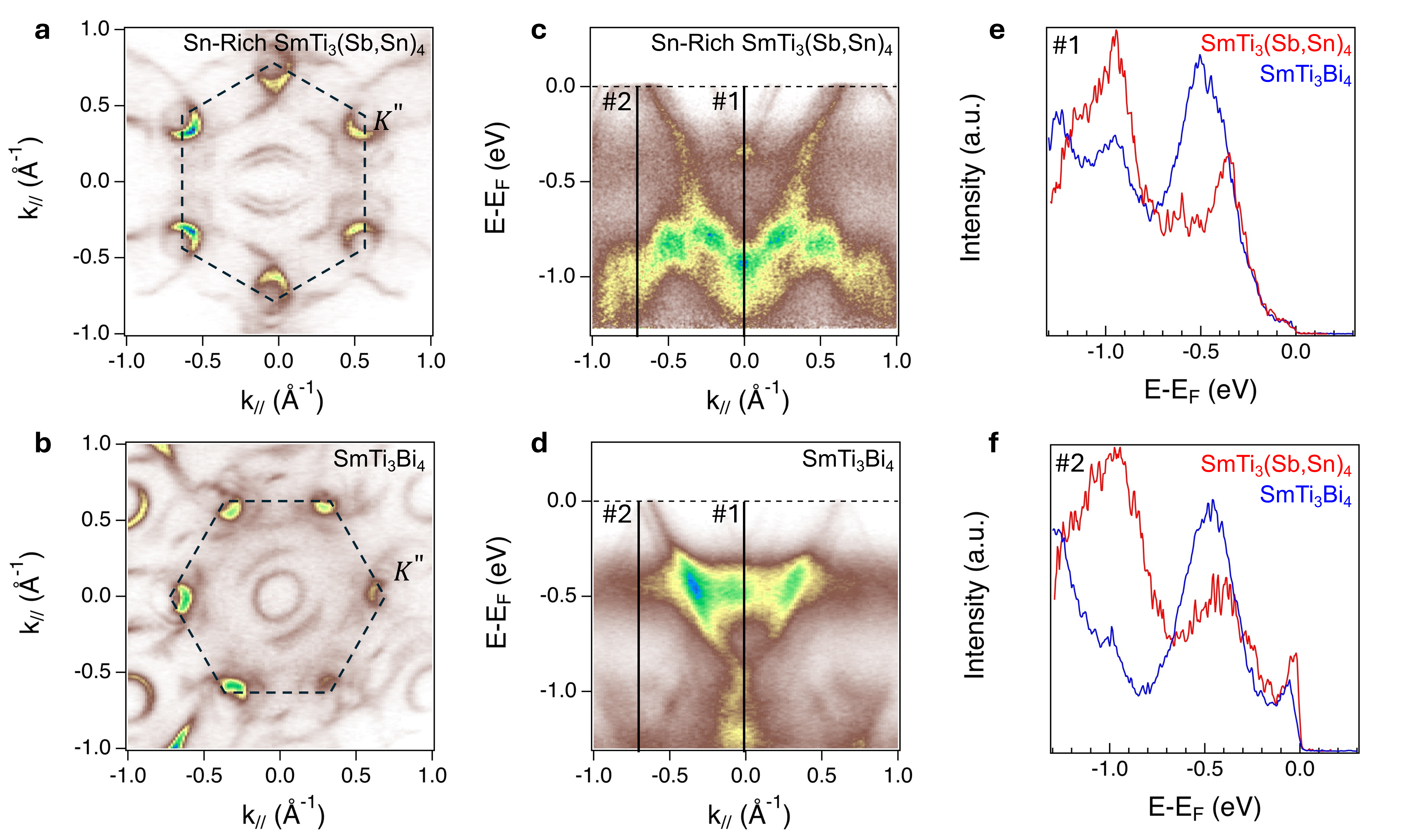}
\caption{Supplementary ARPES data collected on the Sb-rich \SmSS and the bismuthide SmTi$_3$Bi$_4$. (a,b) Constant energy contour (CEC) ARPES data for \SmSS~ and SmTi$_3$Bi$_4$, respectively. Dashed hexagons highlight the Brillouin zone boundary on the surface. (c,d) ARPES intensity plots along K-G-K direction, corresponding to the red dashed line shown in CEC plots. (e) Comparison of the energy distribution curves (EDCs) at the G point (cut 1). Blue and red EDC's represent SmTi$_3$Bi$_4$ and \SmSS, respectively. (f) Comparison of EDC's at the K point (cut 2). Comparing with SmTi$_3$Bi$_4$, the shift in the electronic structure of \SmSS~  is consistent with hole doping in \SmSS. 
}
\label{fig:HMarpes}
\end{figure}
\vspace*{\fill}
\clearpage

\clearpage
\thispagestyle{empty}
\vspace*{\fill}
\begin{figure}[H]
\centering
\includegraphics[width=1\textwidth]{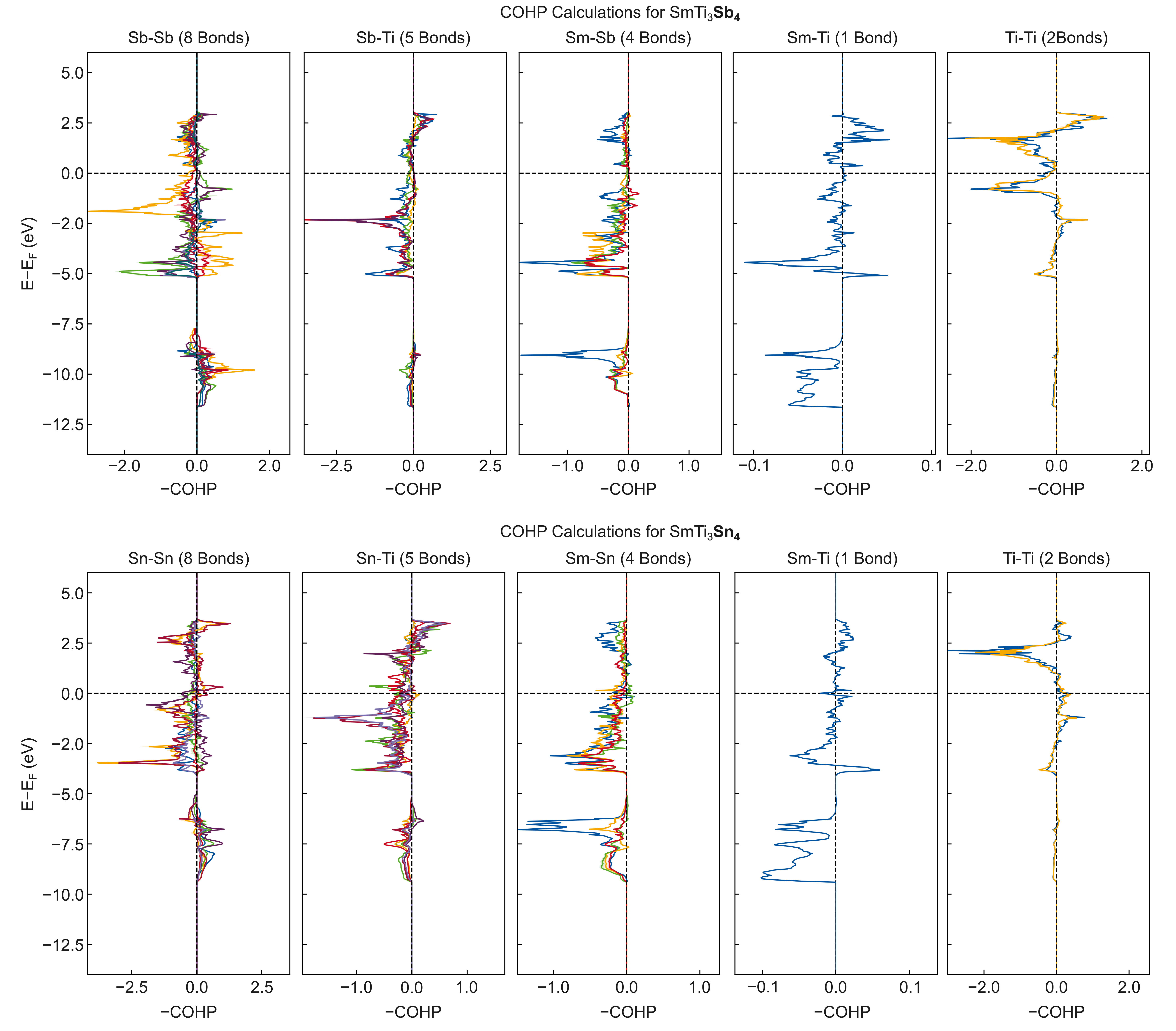}
\caption{Supplementary Crystal Orbital Hamilton Population (COHP) data for both SmTi$_3$Sb$_4$ (top 5 panels) and SmTi$_3$Sn$_4$ (bottom 5 panels). In the main body we focused on the Ti-Ti, Sb-Sb, and Sn-Sn interactions, but here we show all 5 types of interactions.}
\label{fig:COHP}
\end{figure}
\vspace*{\fill}
\clearpage

\clearpage
\thispagestyle{empty}
\vspace*{\fill}
\begin{figure}[H]
\centering
\includegraphics[width=0.8\textwidth]{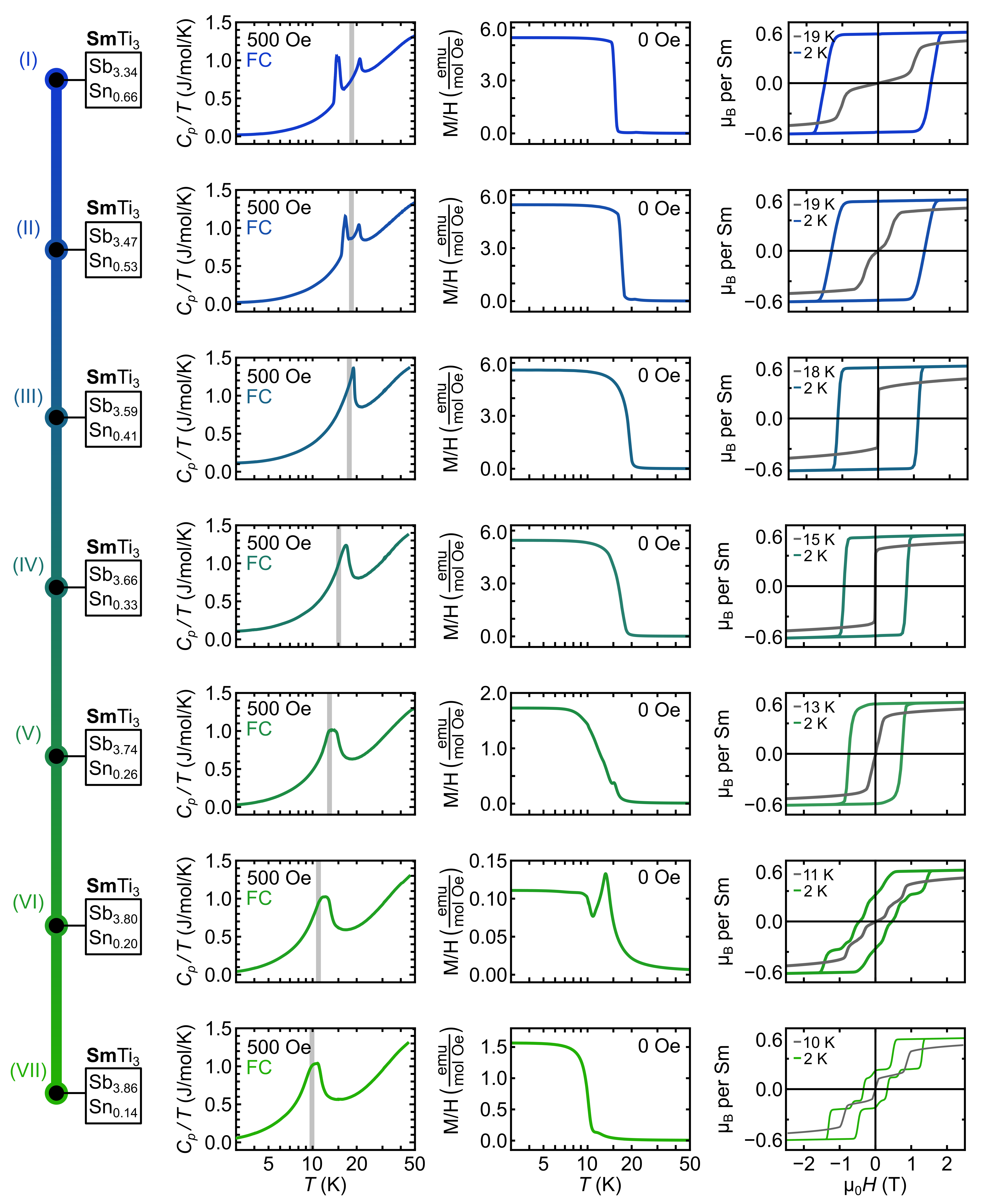}
\caption{Supplementary characterization data for the full series of samples traversing the \SmSS~ solid-solution to demonstrate the gradual shift in magnetic and thermodynamic properties. This is identical data to that shown in the main body, but separated onto individual axes for quantitative comparison. Shown from left to right are: (1) the measured (Sb,Sn) content of the single crystals, (2) a plot showing the temperature-dependent magnetization (3) zero-field heat capacity results, and (4) the corresponding isothermal magnetization curves at base temperature and an intermediate temperature (gray) intended to demarcate the transition between the two heat capacity features (where possible). For clarity, the intermediate temperature (gray) is also marked on the temperature-dependent magnetization.}
\label{fig:Nd}
\end{figure}
\vspace*{\fill}
\clearpage

\clearpage
\thispagestyle{empty}
\vspace*{\fill}
\begin{figure}[H]
\centering
\includegraphics[width=0.6\textwidth]{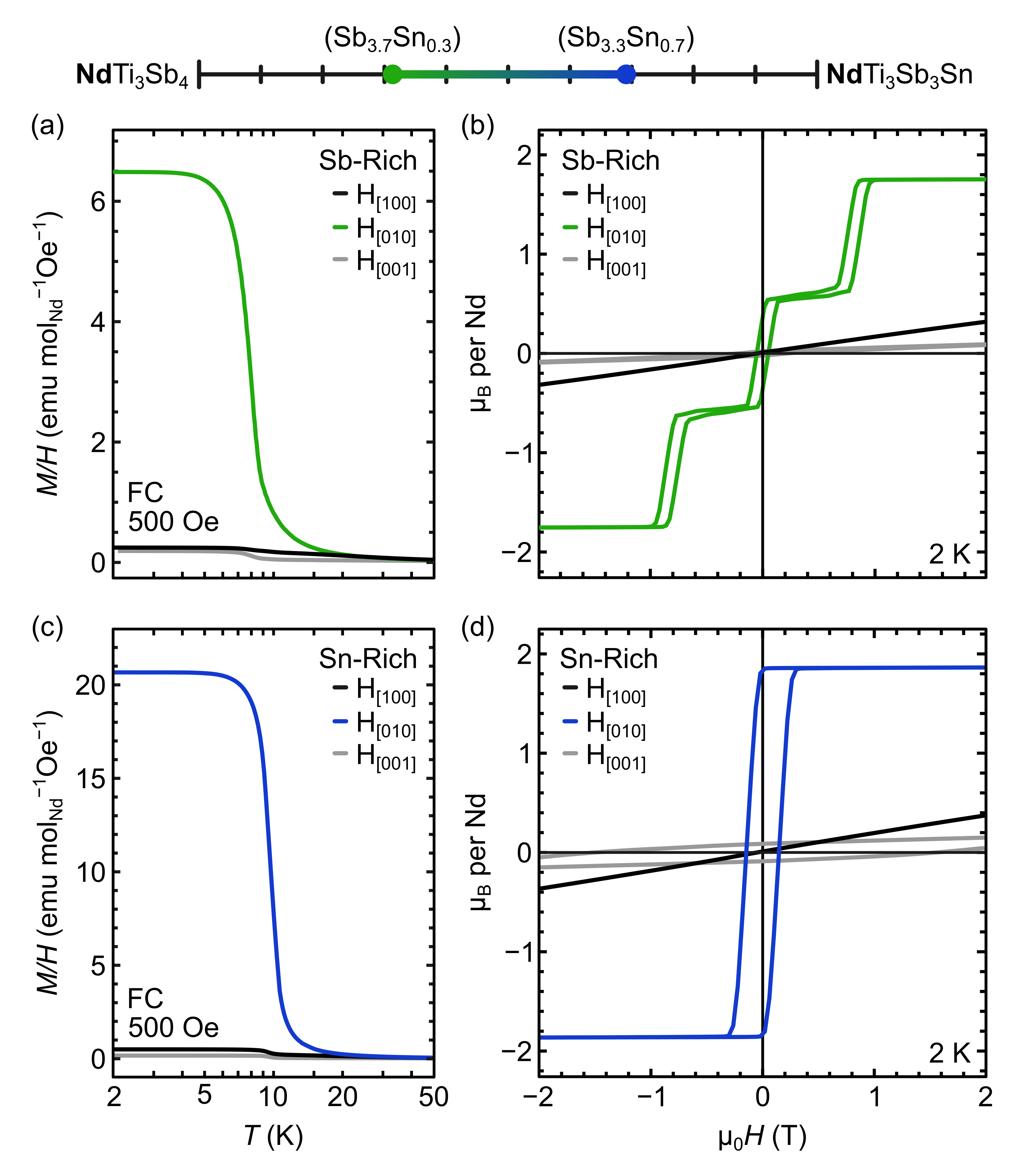}
\caption{Supplementary characterization data for \NdSS~ alloy series. The Nd-containing series shows properties remarkably reminiscent of the \SmSS~ data shown previously. Temperature-dependent magnetization for Sb-rich \NdSS~ (a) shows a broad peak with with a plateau which is weaker than would be expected for a prototypical FM. Isothermal magnetization results (b) reveal field-induced metamagnetic transitions with small but appreciable hysteresis in both transitions, reminiscent of the A(FM) state proposed for Sb-rich \SmSS. Sn-rich \NdSS~ resembles a more prototypical FM with onset of magnetic order around 10~K in temperature-dependent magnetization measurements (c). Isothermal magnetization (d) confirms a FM response with a relatively weak coercive field less than 0.2~T. Note that \NdSS~ exhibits a easy-axis along \textit{b}.}
\label{fig:Nd}
\end{figure}

\vspace*{\fill}
\clearpage

\clearpage
\thispagestyle{empty}
\vspace*{\fill}
\begin{figure}[H]
\centering
\includegraphics[width=0.6\textwidth]{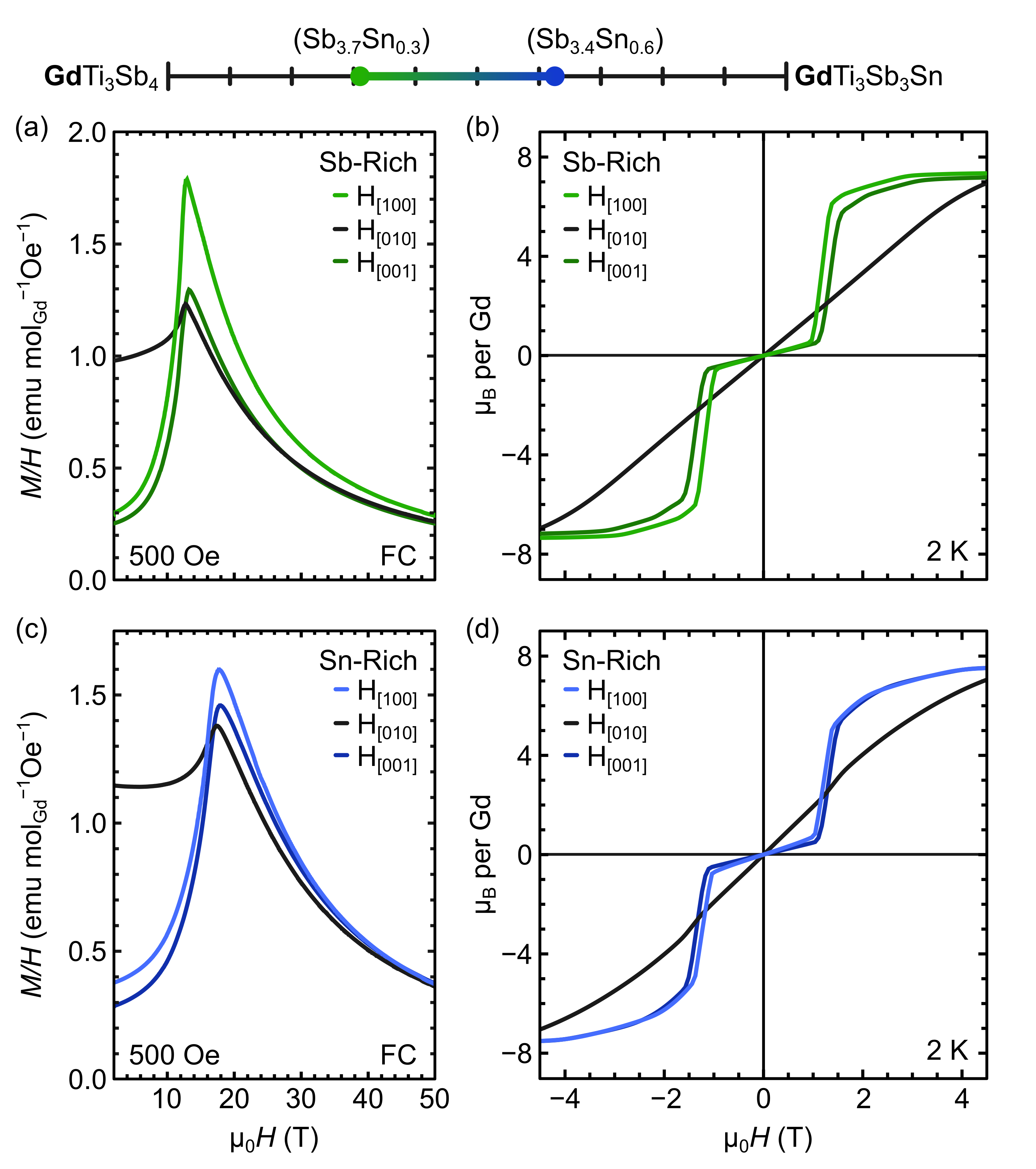}
\caption{Supplementary characterization data for \GdSS~ alloy series. The Gd-containing series shows properties consistent with the bismuthide GdTi$_3$Bi$_4$ published previously. Despite the substantial (Sb,Sn) solubility range in \GdSS, the system is substantially less responsive to the doping, potentially due to the large spin-only moment of Gd$^{3+}$. Both the Sb-rich (a) and Sn-rich (c) temperature-dependent magnetization exhibit onset of AFM order around 13~K and 18~K, respectively. Both isothermal magnetization curves (b,d) are remarkably similar and plateau to values expected for the full Gd$^{3+}$ moment. A clear field-induced spin-flop transition replete with a gradual spin reorientation is observed. Field-induced metamagnetism in the \GdSS~ series occurs for fields directed along either \textit{c} or \textit{a}, consistent with GdTi$_3$Bi$_4$. These results suggest that this compound may have a non-collinear AFM structure or other complex anisotropy.}
\label{fig:Gd}
\end{figure}
\vspace*{\fill}
\clearpage

\clearpage
\thispagestyle{empty}
\vspace*{\fill}
\begin{figure}[H]
\centering
\includegraphics[width=0.6\textwidth]{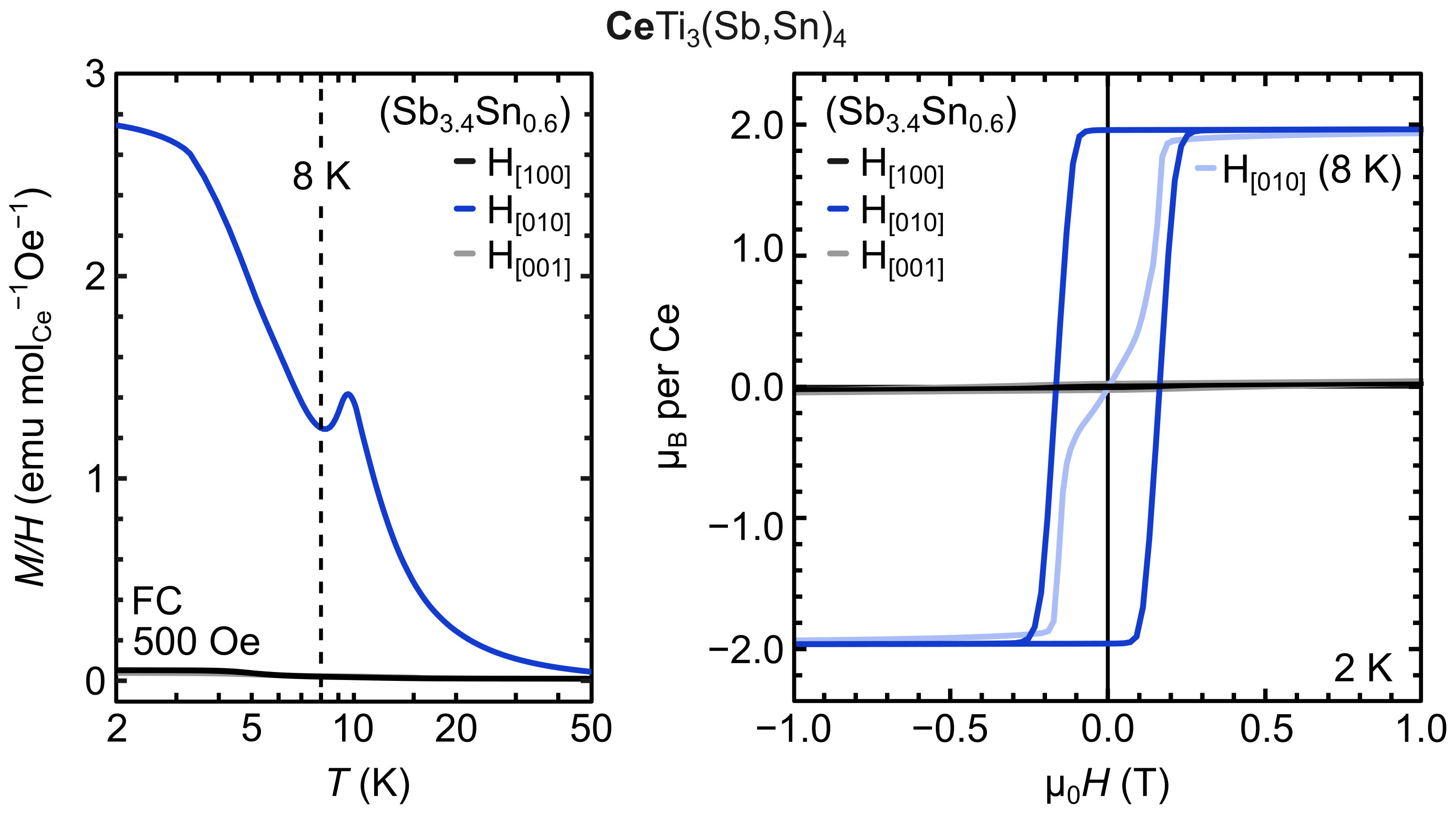}
\caption{Supplementary characterization data for \CeSS~ alloy series. The Ce-containing compound exhibits an AFM cusp at 10~K followed by a FM-like (easy \textit{b}-axis) response that saturates upon cooling below 3~K. Properties are broadly consistent with the magnetization behavior seen in the Sn-rich \SmSS~ series. Field-dependent measurements at 8~K confirm AFM (metamagnetic) nature of the 10~K peak. As in Sn-rich \SmSS, the AFM response appears to be suppressed and replaced with FM order by 2~K.}
\label{fig:Ce}
\end{figure}

\begin{figure}[H]
\centering
\includegraphics[width=0.6\textwidth]{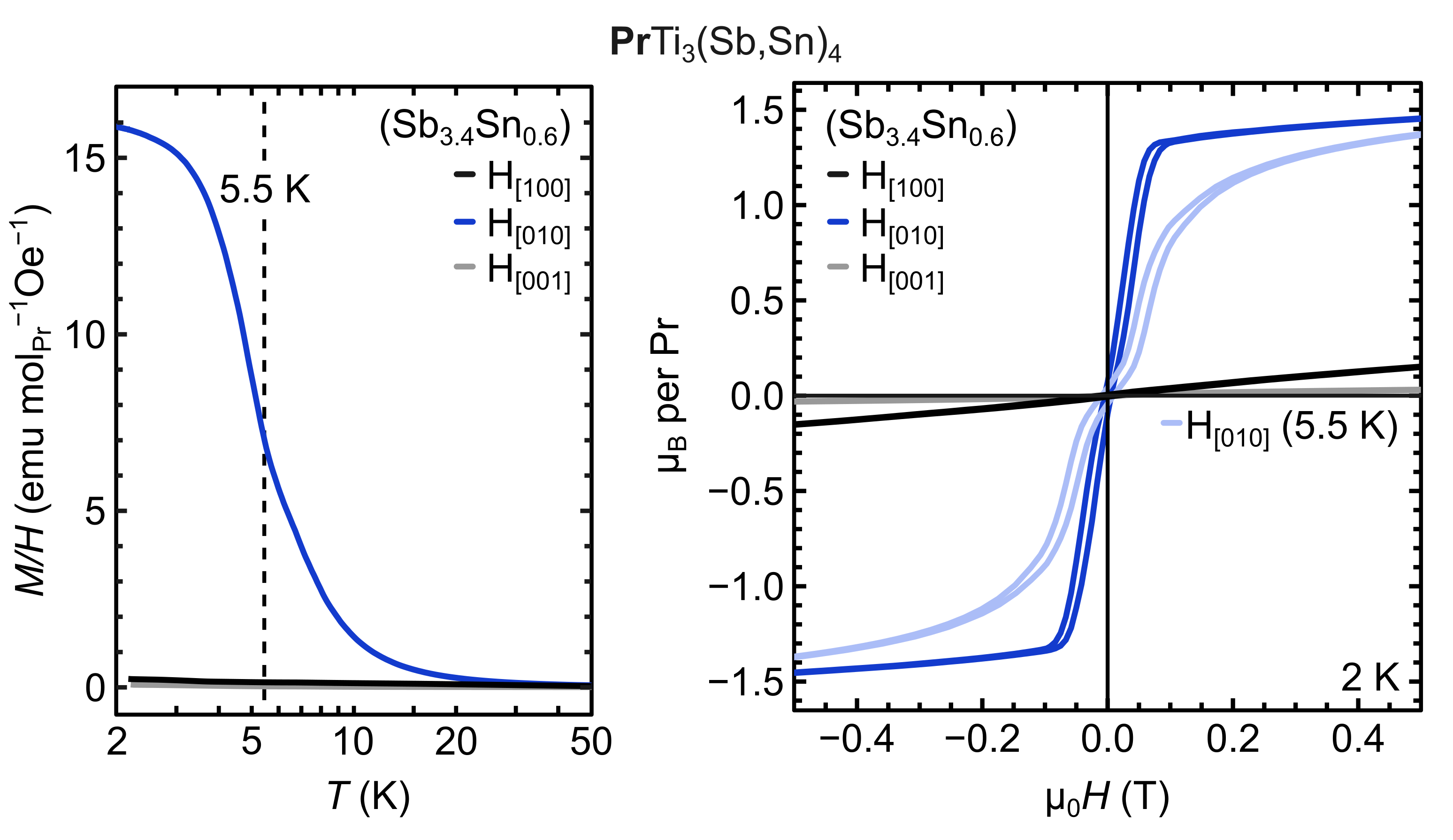}
\caption{Supplementary characterization data for \PrSS~ alloy series. The Pr-containing compound exhibits a broad magnetization rise below approximately 5.5~K. Perhaps \PrSS~ exhibits features similar to the rest of the rare-earth compounds (e.g. a broad AFM shoulder followed by a FM rise with easy \textit{b}-axis magnetization), but the details are not clear. The case of Pr is further complicated by the known sensitivity of the Pr-ion to local disorder and CEF effects.}
\label{fig:Pr}
\end{figure}
\vspace*{\fill}
\clearpage

\clearpage
\thispagestyle{empty}
\vspace*{\fill}
\begin{table*}[htbp]
\centering
\caption{Abbreviated summary of supplementary crystal data, structural refinement, and atomic positions for Sb-rich \SmSS. Specific composition is SmTi$_3$Sb$_{3.86}$Sn$_{0.14}$ (corresponding to green data traces in main body). Complete CIF file is provided in the supplementary information. Specific (Sb,Sn) composition is set by spectroscopic methods and is assumed to be distributed randomly across all Sb sites based on preliminary neutron diffraction data.}
\label{tab:combined}
\begin{tabular}{@{}ll @{\hspace{1cm}} lccccc@{}}
\toprule
\multicolumn{2}{c}{Crystal Data and Refinement} &
\multicolumn{6}{c}{Atomic Positions (fractional coordinates)} \\
\cmidrule(lr){1-2} \cmidrule(lr){3-8}
Property & Value &
Label & $x$ & $y$ & $z$ & $U_{\mathrm{iso}}$ & Occ. \\
\midrule

Formula & \textbf{SmTi$_3$Sb$_{3.86}$Sn$_{0.14}$} &
Sm01 & 1 & 1/2 & 0.69540(2) & 0.00767(10) & 1 \\

Space group, $Z$ & Fmmm, 8 &
Ti01 & 1/2 & 1/2 & 0.59181(5) & 0.0071(2) & 1 \\

$a$ (\AA) & 5.76500(10) &
Ti02 & 1/4 & 1/4 & 0.40441(3) & 0.00664(15) & 1 \\

$b$ (\AA) & 10.0901(2) &
Sb01 & 1/2 & 0.32901(4) & 1/2 & 0.00753(11) & 0.9647 \\

$c$ (\AA) & 24.0655(5) &
Sn01 & '' & '' & '' & '' & 0.0353 \\

Volume (\AA$^3$) & 1399.88(5) &
Sb02 & 0 & 0.15948(3) & 0.31149(2) & 0.00736(9) & 0.9647 \\

Density (calc.) (g cm$^{-3}$) & 7.408 &
Sn02 & '' & '' & '' & '' & 0.0353 \\

$R_{\mathrm{int}}$, Completeness & 0.0230, 99\% &
Sb03 & 1 & 1/2 & 0.56641(2) & 0.00731(11) & 0.9647 \\

$R_1$ [$I>2\sigma(I)$] & 0.0226 &
Sn03 & '' & '' & '' & '' & 0.0353 \\

$wR_2$ [$I>2\sigma(I)$] & 0.0677 & & & & & & \\

Largest peak/hole (e \AA$^{-3}$) & 3.291 / $-2.737$ & & & & & & \\

\bottomrule
\end{tabular}
\end{table*}

\begin{table*}[htbp]
\centering
\caption{Abbreviated summary of supplementary crystal data, structural refinement, and atomic positions for Sn-rich \SmSS. Specific composition is SmTi$_{3}$Sb$_{3.338}$Sn$_{0.662}$ (corresponding to blue data traces in main body). Complete CIF file is provided in the supplementary information. Specific (Sb,Sn) composition is set by spectroscopic methods and is assumed to be distributed randomly across all Sb sites based on preliminary neutron diffraction data.}
\label{tab:combined}
\begin{tabular}{@{}ll @{\hspace{1cm}} lccccc@{}}
\toprule
\multicolumn{2}{c}{Crystal Data and Refinement} &
\multicolumn{6}{c}{Atomic Positions (fractional coordinates)} \\
\cmidrule(lr){1-2} \cmidrule(lr){3-8}
Property & Value &
Label & $x$ & $y$ & $z$ & $U_{\mathrm{iso}}$ & Occ. \\
\midrule

Formula & \textbf{SmTi$_{3}$Sb$_{3.338}$Sn$_{0.662}$} &
Sm01 & 1 & 1/2 & 0.69508(2) & 0.0075(2) & 1 \\

Space group, $Z$ & Fmmm, 8 &
Ti01 & 1/2 & 1/2 & 0.59269(9) & 0.0077(4) & 1 \\

$a$ (\AA) & 5.7582(3) &
Ti02 & 1/4 & 1/4 & 0.40362(6) & 0.0068(3) & 1 \\

$b$ (\AA) & 10.0696(5) &
Sb01 & 1/2 & 0.32890(8) & 1/2 & 0.0088(2) & 0.8345 \\

$c$ (\AA) & 24.1350(11) &
Sn01 & '' & '' & '' & '' & 0.1655 \\

Volume (\AA$^3$) & 1399.41(12) &
Sb02 & 0 & 0.16014(5) & 0.31100(2) & 0.0073(2) & 0.8345 \\

Density (calc.) (g cm$^{-3}$) & 7.395 &
Sn02 & '' & '' & '' & '' & 0.1655 \\

$R_{\mathrm{int}}$, Completeness & 0.0344, 98.3\% &
Sb03 & 1 & 1/2 & 0.56622(3) & 0.0077(2) & 0.8345 \\

$R_1$ [$I>2\sigma(I)$] & 0.0335 &
Sn03 & '' & '' & '' & '' & 0.1655 \\

$wR_2$ [$I>2\sigma(I)$] & 0.0838 & & & & & & \\

Largest peak/hole (e \AA$^{-3}$) & 3.703 / $-2.649$ & & & & & & \\

\bottomrule
\end{tabular}
\end{table*}
\vspace*{\fill}
\clearpage

\clearpage
\thispagestyle{empty}
\vspace*{\fill}

\begin{table*}[htbp]
\centering
\caption{Abbreviated summary of supplementary crystal data, structural refinement, and atomic positions for Sb-rich \CeSS. Specific composition is CeTi$_{3}$Sb$_{3.38}$Sn$_{0.62}$ (corresponding to green data traces in main body). Complete CIF file is provided in the supplementary information. Specific (Sb,Sn) composition is set by spectroscopic methods and is assumed to be distributed randomly across all Sb sites based on preliminary neutron diffraction data.}
\label{tab:combined}
\begin{tabular}{@{}ll @{\hspace{1cm}} lccccc@{}}
\toprule
\multicolumn{2}{c}{Crystal Data and Refinement} &
\multicolumn{6}{c}{Atomic Positions (fractional coordinates)} \\
\cmidrule(lr){1-2} \cmidrule(lr){3-8}
Property & Value &
Label & $x$ & $y$ & $z$ & $U_{\mathrm{iso}}$ & Occ. \\
\midrule

Formula & \textbf{CeTi$_{3}$Sb$_{3.38}$Sn$_{0.62}$} &
Ce01 & 1 & 1/2 & 0.69513(4) & 0.0099(2) & 1 \\

Space group, $Z$ & Fmmm, 8 &
Ti01 & 1/2 & 1/2 & 0.59171(15) & 0.0103(6) & 1 \\

$a$ (\AA) & 5.8102(8) &
Ti02 & 1/4 & 1/4 & 0.40523(9) & 0.0086(4) & 1 \\

$b$ (\AA) & 10.1127(13) &
Sb01 & 1/2 & 0.32947(14) & 1/2 & 0.0120(3) & 0.845 \\

$c$ (\AA) & 24.402(3) &
Sn01 & '' & '' & '' & '' & 0.155 \\

Volume (\AA$^3$) & 1433.8(3) &
Sb02 & 0 & 0.16112(9) & 0.31356(4) & 0.0097(2) & 0.845 \\

Density (calc.) (g cm$^{-3}$) & 7.124 &
Sn02 & '' & '' & '' & '' & 0.155 \\

$R_{\mathrm{int}}$, Completeness & 0.0888, 99.3\% &
Sb03 & 1 & 1/2 & 0.56523(5) & 0.0103(3) & 0.8455 \\

$R_1$ [$I>2\sigma(I)$] & 0.0543 &
Sn03 & '' & '' & '' & '' & 0.155 \\

$wR_2$ [$I>2\sigma(I)$] & 0.1143 & & & & & & \\

Largest peak/hole (e \AA$^{-3}$) & 4.526 / $-3.423$ & & & & & & \\

\bottomrule
\end{tabular}
\end{table*}

\begin{table*}[htbp]
\centering
\caption{Abbreviated summary of supplementary crystal data, structural refinement, and atomic positions for Sb-rich \PrSS. Specific composition is PrTi$_{3}$Sb$_{3.365}$Sn$_{0.635}$ (corresponding to green data traces in main body). Complete CIF file is provided in the supplementary information. Specific (Sb,Sn) composition is set by spectroscopic methods and is assumed to be distributed randomly across all Sb sites based on preliminary neutron diffraction data.}
\label{tab:combined}
\begin{tabular}{@{}ll @{\hspace{1cm}} lccccc@{}}
\toprule
\multicolumn{2}{c}{Crystal Data and Refinement} &
\multicolumn{6}{c}{Atomic Positions (fractional coordinates)} \\
\cmidrule(lr){1-2} \cmidrule(lr){3-8}
Property & Value &
Label & $x$ & $y$ & $z$ & $U_{\mathrm{iso}}$ & Occ. \\
\midrule

Formula & \textbf{PrTi$_{3}$Sb$_{3.365}$Sn$_{0.635}$} &
Pr01 & 1 & 1/2 & 0.69503(2) & 0.00792(10) & 1 \\

Space group, $Z$ & Fmmm, 8 &
Ti01 & 1/2 & 1/2 & 0.59221(6) & 0.0080(2) & 1 \\

$a$ (\AA) & 5.7999(2) &
Ti02 & 1/4 & 1/4 & 0.40457(4) & 0.00728(16) & 1 \\

$b$ (\AA) & 10.1034(4) &
Sb01 & 1/2 & 0.32929(5) & 1/2 & 0.00967(11) & 0.8412 \\

$c$ (\AA) & 24.3298(9) &
Sn01 & '' & '' & '' & '' & 0.1588 \\

Volume (\AA$^3$) & 1425.69(9) &
Sb02 & 0 & 0.16096(4) & 0.31269(2) & 0.00796(9) & 0.8412 \\

Density (calc.) (g cm$^{-3}$) & 7.172 &
Sn02 & '' & '' & '' & '' & 0.1587 \\

$R_{\mathrm{int}}$, Completeness & 0.0302, 99.8\% &
Sb03 & 1 & 1/2 & 0.56531(2) & 0.00842(11) & 0.8412 \\

$R_1$ [$I>2\sigma(I)$] & 0.0286 &
Sn03 & '' & '' & '' & '' & 0.1588 \\

$wR_2$ [$I>2\sigma(I)$] & 0.0666 & & & & & & \\

Largest peak/hole (e \AA$^{-3}$) & 3.024 / $-1.751$ & & & & & & \\

\bottomrule
\end{tabular}
\end{table*}
\vspace*{\fill}
\clearpage

\clearpage
\thispagestyle{empty}
\vspace*{\fill}
\begin{table*}[htbp]
\centering
\caption{Abbreviated summary of supplementary crystal data, structural refinement, and atomic positions for Sb-rich \NdSS. Specific composition is NdTi$_{3}$Sb$_{3.29}$Sn$_{0.71}$ (corresponding to green data traces in main body). Complete CIF file is provided in the supplementary information. Specific (Sb,Sn) composition is set by spectroscopic methods and is assumed to be distributed randomly across all Sb sites based on preliminary neutron diffraction data.}
\label{tab:combined}
\begin{tabular}{@{}ll @{\hspace{1cm}} lccccc@{}}
\toprule
\multicolumn{2}{c}{Crystal Data and Refinement} &
\multicolumn{6}{c}{Atomic Positions (fractional coordinates)} \\
\cmidrule(lr){1-2} \cmidrule(lr){3-8}
Property & Value &
Label & $x$ & $y$ & $z$ & $U_{\mathrm{iso}}$ & Occ. \\
\midrule

Formula & \textbf{NdTi$_{3}$Sb$_{3.29}$Sn$_{0.71}$} &
Nd01 & 1 & 1/2 & 0.69504(2) & 0.00780(9) & 1 \\

Space group, $Z$ & Fmmm, 8 &
Ti01 & 1/2 & 1/2 & 0.59238(5) & 0.0076(2) & 1 \\

$a$ (\AA) & 5.7839(3) &
Ti02 & 1/4 & 1/4 & 0.40421(3) & 0.00694(14) & 1 \\

$b$ (\AA) & 10.0918(5) &
Sb01 & 1/2 & 0.32910(5) & 1/2 & 0.00923(10) & 0.8225 \\

$c$ (\AA) & 24.2738(13) &
Sn01 & '' & '' & '' & '' & 0.1775 \\

Volume (\AA$^3$) & 1416.86(13) &
Sb02 & 0 & 0.16062(3) & 0.31207(2) & 0.00762(9) & 0.8225 \\

Density (calc.) (g cm$^{-3}$) & 7.245 &
Sn02 & '' & '' & '' & '' & 0.1775 \\

$R_{\mathrm{int}}$, Completeness & 0.0263, 99.1\% &
Sb03 & 1 & 1/2 & 0.56559(2) & 0.00793(10) & 0.8225 \\

$R_1$ [$I>2\sigma(I)$] & 0.0257 &
Sn03 & '' & '' & '' & '' & 0.1775 \\

$wR_2$ [$I>2\sigma(I)$] & 0.0608 & & & & & & \\

Largest peak/hole (e \AA$^{-3}$) & 1.747 / $-2.495$ & & & & & & \\

\bottomrule
\end{tabular}
\end{table*}

\begin{table*}[htbp]
\centering
\caption{Abbreviated summary of supplementary crystal data, structural refinement, and atomic positions for Sb-rich \GdSS. Specific composition is GdTi$_{3}$Sb$_{3.373}$Sn$_{0.627}$ (corresponding to green data traces in main body). Complete CIF file is provided in the supplementary information. Specific (Sb,Sn) composition is set by spectroscopic methods and is assumed to be distributed randomly across all Sb sites based on preliminary neutron diffraction data.}
\label{tab:combined}
\begin{tabular}{@{}ll @{\hspace{1cm}} lccccc@{}}
\toprule
\multicolumn{2}{c}{Crystal Data and Refinement} &
\multicolumn{6}{c}{Atomic Positions (fractional coordinates)} \\
\cmidrule(lr){1-2} \cmidrule(lr){3-8}
Property & Value &
Label & $x$ & $y$ & $z$ & $U_{\mathrm{iso}}$ & Occ. \\
\midrule

Formula & \textbf{GdTi$_{3}$Sb$_{3.373}$Sn$_{0.627}$} &
Gd01 & 1 & 1/2 & 0.69527(2) & 0.00765(9) & 1 \\

Space group, $Z$ & Fmmm, 8 &
Ti01 & 1/2 & 1/2 & 0.59274(5) & 0.0066(2) & 1 \\

$a$ (\AA) & 5.7467(3) &
Ti02 & 1/4 & 1/4 & 0.40345(4) & 0.00650(16) & 1 \\

$b$ (\AA) & 10.0645(6) &
Sb01 & 1/2 & 0.32888(5) & 1/2 & 0.00840(11) & 0.8432 \\

$c$ (\AA) & 24.0610(14) &
Sn01 & '' & '' & '' & '' & 0.1568 \\

Volume (\AA$^3$) & 1391.63(14) &
Sb02 & 0 & 0.15952(3) & 0.31043(2) & 0.00728(9) & 0.8433 \\

Density (calc.) (g cm$^{-3}$) & 7.503 &
Sn02 & '' & '' & '' & '' & 0.1567 \\

$R_{\mathrm{int}}$, Completeness & 0.0286, 99.8\% &
Sb03 & 1 & 1/2 & 0.56680(2) & 0.00729(11) & 0.8432 \\

$R_1$ [$I>2\sigma(I)$] & 0.0270 &
Sn03 & '' & '' & '' & '' & 0.1568 \\

$wR_2$ [$I>2\sigma(I)$] & 0.0662 & & & & & & \\

Largest peak/hole (e \AA$^{-3}$) & 2.545 / $-3.031$ & & & & & & \\

\bottomrule
\end{tabular}
\end{table*}
\vspace*{\fill}
\clearpage

\clearpage
\thispagestyle{empty}
\vspace*{\fill}
\begin{figure}[H]
\centering
\includegraphics[width=0.7\textwidth]{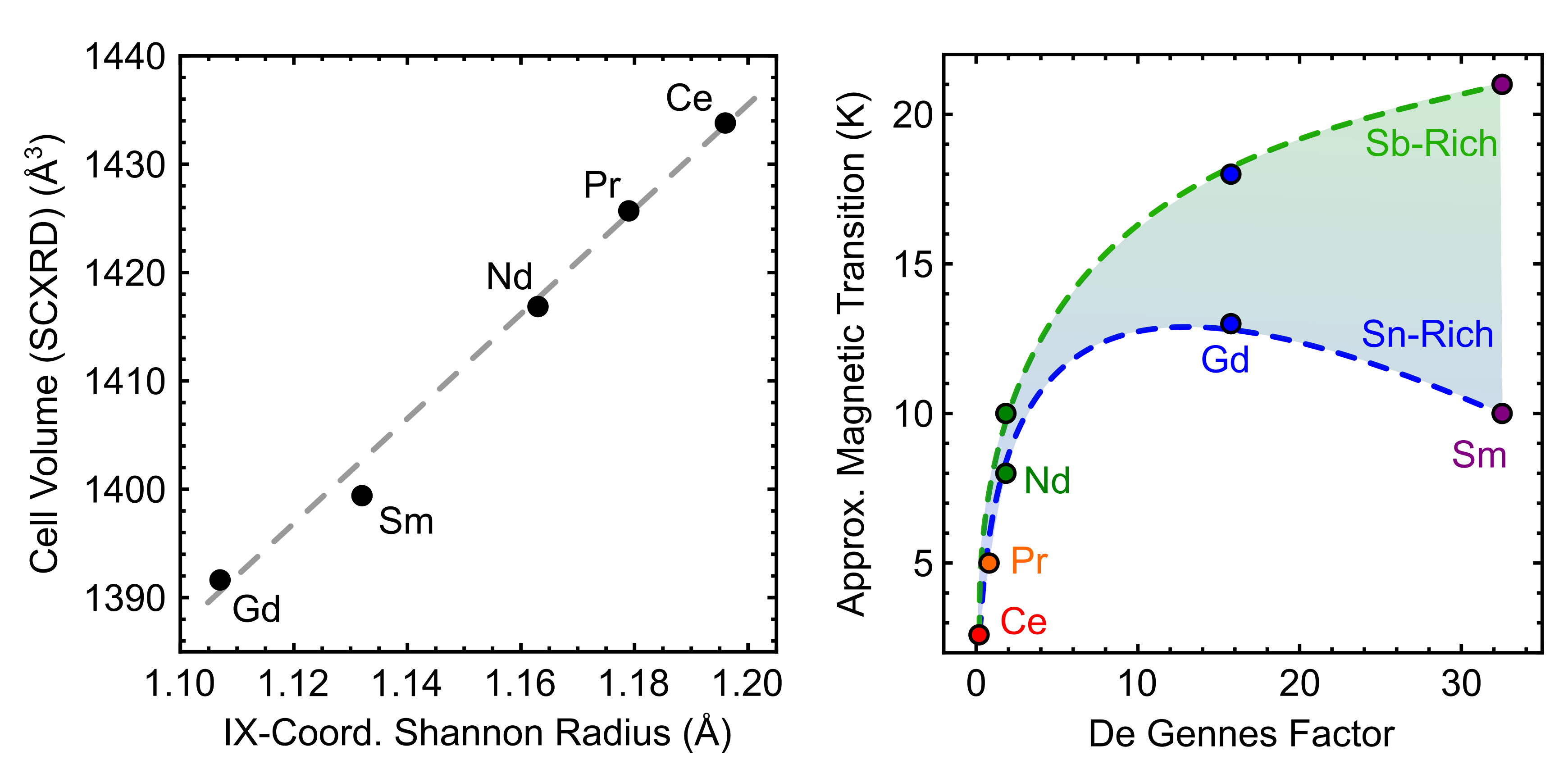}
\caption{Supplementary summary plots showing (a) the trend with lattice parameter throughout the rare-earth series, and (b) the De Gennes factor versus the magnetic transition temperatures, where both the Sn-rich and Sb-rich compound are shown for each rare-earth.}
\label{fig:Pr}
\end{figure}
\vspace*{\fill}
\clearpage

\end{document}